\documentclass[preprint,12pt]{elsarticle}

\usepackage{amssymb}
\usepackage{longtable}
\usepackage{url}
\usepackage{amsmath}

\usepackage{booktabs}
\usepackage{cite}

\begin{document}

\makeatletter
\def\ps@pprintTitle{%
  \let\@oddhead\@empty
  \let\@evenhead\@empty
  \let\@oddfoot\@empty
  \let\@evenfoot\@oddfoot
}
\makeatother

\begin{frontmatter}



\title{Power System Transition Planning: An Industry-Aligned Framework for Long-Term Optimization} 




\author[inst1]{Ahmed Al-Shafei\corref{cor1}\fnref{fn1}}
\ead{ahmed.alshafei@ucalgary.ca}

\author[inst2]{Nima Amjady}
\author[inst1]{Hamidreza Zareipour}
\author[inst3]{Yankai Cao}

\cortext[cor1]{Corresponding author.}

\affiliation[inst1]{organization={Department
of Electrical and Computer Engineering, University of Calgary}, 
            addressline={2500 University Dr NW}, 
            city={Calgary}, 
            postcode={T2N 1N4}, 
            state={AB}, 
            country={Canada}}

\affiliation[inst2]{organization={Centre for New Energy Transition Research (CfNETR), Federation University}, 
            addressline={MT Campus}, 
            city={Ballarat}, 
            postcode={3350}, 
            state={VIC}, 
            country={Australia}}

\affiliation[inst3]{organization={Department of Chemical and Biological Engineering, University of British Columbia}, 
            addressline={2360 East Mall}, 
            city={Vancouver}, 
            postcode={V6T 1Z3}, 
            state={BC}, 
            country={Canada}}


\begin{abstract}
This work introduces the category of Power System Transition Planning optimization problem. It aims to shift power systems to emissions-free networks efficiently. Unlike comparable work, the framework presented here broadly applies to the industry's decision-making process. It defines a field-appropriate functional boundary focused on the economic efficiency of power systems. Namely, while imposing a wide range of planning factors in the decision space, the model maintains the structure and depth of conventional power system planning under uncertainty, which leads to a large-scale multistage stochastic programming formulation that encounters intractability in real-life cases. Thus, the framework simultaneously invokes high-performance computing defaultism. In this comprehensive exposition, we present a guideline model, comparing its scope to existing formulations, supported by a fully detailed example problem, showcasing the analytical value of the solution gained in a small test case. Then, the framework's viability for realistic applications is demonstrated by solving an extensive test case based on a realistic planning construct consistent with Alberta's power system practices for long-term planning studies. The framework resorts to Stochastic Dual Dynamic Programming as a decomposition method to achieve tractability, leveraging High-Performance Computing and parallel computation.
\end{abstract}





\end{frontmatter}



\section*{Nomenclature}

\renewcommand{\arraystretch}{1.2}
\begin{longtable}{p{0.25\textwidth} p{0.75\textwidth}}
\multicolumn{2}{l}{\textbf{Indices and Sets}} \\
$g, \Omega_G$               & Index and set of existing generation units \\
$l, \Omega_{R}$             & Index and set of Rights-of-Way (RoW) \\
$s, \Omega_{S}$             & Index and set of scenarios \\
$t, \Omega_{T}$             & Index and set of operational hours \\
$n, \Omega_{B}$             & Index and set of substations \\
$o, \Omega_{O}$             & Index and set of operational conditions \\
$y, \Omega_{Y}$             & Index and set of decision stages \\
$z$                         & Index of candidate renewable energy zones \\
$\Omega_{Zs}, \Omega_{Zw}$  & Sets of solar and wind candidate zones \\
$\Omega_{Zs,n}$             & Solar zones associated with bus $n$ \\
$\Omega_{Zw,n}$             & Wind zones associated with bus $n$ \\
$\tau$                      & Stage summation index \\
\\

\multicolumn{2}{l}{\textbf{Parameters}} \\
$\eta^{CH}$, $\eta^{DI}$    & Charging and discharging efficiencies \\
$\gamma$                    & Asset lifetime [years] \\
$\phi_s$                    & Probability of scenario $s$ \\
$\rho_o$                    & Weight of operational condition $o$ \\
$\sigma^T$, $\sigma^P$      & Water flow to power conversion factors \\
$\theta^{\max}, \theta^{\min}$ & Max/min bus angle \\
$A$                         & Available area limit [$km^2$] \\
$C^{in}$                    & SSSC cut-in current \\
$CH^{\max}$, $DI^{\max}$    & Max charging/discharging rate [$MW/h$] \\
$CO_2$                      & Generator CO\textsubscript{2} emission [$tonnes/MW$] \\
$G_n$                       & Max new connection at bus $n$ [$MW$] \\
$F^{\max}_l$, $F^{\min}_l$  & Max/min SSSC operational factor on line $l$ \\
$I$                         & Technology cost [$\$$ or \$/MW] \\
$K$                         & Land use per MW [$km^2$/MW] \\
$N_l$                       & Existing line indicator in RoW $l$ \\
$P$                         & Variable cost [$\$/MW$] \\
$P^{\max}_{\min}$           & Max/min existing generator output [$MW$] \\
$G^{\max}_{\min}$           & Max/min new generator output [$MW$] \\
$Q^{\max}$                  & Max water flow [$Hm^3/h$] \\
$R$                         & Existing capacity [$MW$] \\
$S^{ST,N}_l$, $S^{ST,E}_l$  & Static thermal ratings (new/existing lines) [$MW$] \\
$S^{DTR,N}_l$, $S^{DTR,E}_l$& Dynamic thermal ratings (new/existing lines) [$MW$] \\
$SOC^{\max}_{\min}$         & Max/min state of charge \\
$T_y$                       & CO\textsubscript{2} target in year $y$ [$tonnes$] \\
$UP$, $DN$                  & Ramp-up and ramp-down limits [$MW/h$] \\
$UF$, $DF$                  & Ramp-up/down rate factors [$1/h$] \\
$V^{U,0}, V^{L,0}$          & Initial upper/lower reservoir volume [$Hm^3$] \\
$V^{\max}_{\min}$           & Max/min reservoir volume [$Hm^3$] \\
$X_l$                       & Line reactance in RoW $l$ \\
\\

\multicolumn{2}{l}{\textbf{State Variables}} \\
$i$                         & Installed capacity decision [$MW$] \\
$x$                         & Transmission/storage binary/integer decision \\
\\

\multicolumn{2}{l}{\textbf{Random Variables}} \\
$\zeta^W$, $\zeta^S$        & Wind and solar production factors \\
$\zeta^{DTR}$               & Dynamic line rating [$MW$] \\
$\zeta^D_n$                 & Demand at bus $n$ [$MW$] \\
\\

\multicolumn{2}{l}{\textbf{Operational Variables}} \\
$f_l$                       & Power flow in line $l$ [$MW$] \\
$p$                         & Power output or curtailment [$MW$] \\
$r$                         & Reservoir water level [$Hm^3$] \\
$s$                         & State of charge [$MWh$] \\
$\theta$                    & Bus voltage angle [$^\circ$] \\
$v$                         & Water volume [$Hm^3$] \\
$w$                         & Water flow [$Hm^3/h$] \\
$x^{state}$                 & Battery charging state $\{0,1\}$ \\
\end{longtable}
\addtocounter{table}{-1}

\section{Introduction}
\label{introduction}

The global transition to net-zero emissions has dramatically redefined the objectives and challenges of long-term planning of the power system \citep{IEA2021}. Traditional planning methodologies and theoretical models in the literature are increasingly misaligned with the complexity and uncertainty required for a decarbonized future. Most models focus narrowly on deterministic forecasts and isolated frameworks of network expansions that do not capture the breadth of available solutions or account for spatial, temporal, and operational uncertainties inherent in renewable-dominated systems. This work recognizes this need and proposes the Power System Transition Planning (PTSP) category. This comprehensive, geospatially aware planning framework aligns with established industry practices such as those from the Alberta Electric System Operator (AESO) and the North American Electric Reliability Corporation (NERC) and others (see  \ref{app:tables}, Table \ref{tab:PlanPeriod}). These organizations emphasize scenario-based, probabilistic modeling to secure future power delivery under diverse futures, rather than reliance on fixed-point forecasts or purely theoretical models. PTSP builds on this foundation to propose a scalable, uncertainty-aware planning formulation that is tractable for large, real-world systems. 

Industry-standard network-based expansion tools such as PLEXOS, TYNDP, ReEDS, PyPSA, and others (see \ref{app:tables}, Table \ref{tab:software-tools}) are alligned with regulatory and policy frameworks (e.g., TYNDP by ENTSO-E, NERC in North America), which explicitly mandate or encourage scenario-based approaches, embedding scenario planning deeply into policy and regulatory processes. However, they are built around deterministic scenario-based methods, and while deterministic approaches are pragmatic, they generally fail to represent adaptive recourse decisions, multi-stage planning horizons, or uncertainty propagation.

Academic research offers more advanced frameworks for uncertainty modeling, such as Robust Optimization approaches \citep{moreira.etal_2017}, Adaptive robust optimization methods \citep{baringo.baringo_2018}, and hybrid Stochastic-Robust optimization approaches \citep{yin2022a}. Robust optimization approaches offer worst-case performance guarantees, but often produce overly conservative, less intuitive results, making them difficult to integrate into official decision-making processes. In contrast, Stochastic Programming (SP) explicitly models a range of plausible future outcomes, aligning more naturally with scenario-based planning. It supports adaptive recourse actions and enhances decision transparency and interoperability for industry planners \citep{conejo2010decision}.  Multistage stochastic programming (MSP), in particular, allows uncertainty to unfold over time, enabling dynamic, sequential planning. MSP aligns well with strategic scenario modeling in long-term planning reports (e.g., NREL ATB, IPCC)\citep{NRELATB, IPCC2021}. However, practical implementation remains limited due to scalability challenges. Its use has been limited to small systems (e.g., 6–24 buses, 3–6 scenarios)\citep{hou2021} \citep{wei2022} due to scalability challenges. Most MSP applications either simplify the physical network, limit technology breadth, or sacrifice spatial and temporal resolution \citep{yin2022a}.


\subsection{Transition Planning}

The Power System Transition Planning (PSTP) problem is designed to bridge the gap between industrial practicality and academic rigor, and the following pillars define it. The first is that transitioning decisions must occur over several time stages, an aspect often missed in the literature, in which static models dominate. The second is the universality and geographical endowment of the included elements and framework design, meaning that the final design must consider geospatial factors and input. The third is the presence of scenario-based uncertainty, aligning with industry practice. Long-term and short-term randomness has to occur over the decision-making stages. The fourth is the breadth of “planning factors” and technologies involved in the policy-making process, which must be pronounced and include several modular factors and resources that maximize the utilization of the existing grid without resorting to additional infrastructure. Planning Factors are the transition network elements or technologies made available to the planner in policy-making. Their modeling must be sufficiently detailed to ensure its impact is notably reflected in the policy outcome. Finally, the outcome of PSTP should provide the lowest-cost transitioning path alongside information about several other future potential paths that always lead to a low or zero-emission network for policymakers. The critical requirement is not immediate computation but the ability to generate robust, actionable, and defensible plans.

\begin{table*}[b!]
  \caption{Comparative Review of Analogous Studies}
  \label{tab:LitCompare}
  \centering
  \Large
\resizebox{\textwidth}{!}{%
\begin{tabular}{lcccccccccccccccccccccccccccc|c}
    \toprule
        \textbf{Planning Factors} & \citep{wei2022} & \citep{yin2022a} & \citep{2022} & \citep{zhao2022abc} & \citep{hou2021} & \citep{flores-quiroz2021} & \citep{lugovoy2021} & \citep{dacosta2021c} & \citep{shen2020a} & \citep{ramirez2020} & \citep{asgharian2020} & \citep{gbadamosiMultiperiodCompositeGeneration2020} & \citep{alanazi2020a} & \citep{zhou2020} & \citep{asadimajd2020} & \citep{zeinaddini-meymandDemandSideManagementBasedModel2019} & \citep{zhangCoordinationGenerationTransmission2019} & \citep{PARZEN2023121096} & \citep{lara.etal_2020} & \citep{hou.etal_2021} &  This Work   \\ \midrule
        Thermal/Rotary Generation & \checkmark & \checkmark & \checkmark & \checkmark & \checkmark & \checkmark & \checkmark & \checkmark & \checkmark & \checkmark & \checkmark & \checkmark & \checkmark & \checkmark & \checkmark & \checkmark & \checkmark & \checkmark & \checkmark & \checkmark &\checkmark   \\ 
        Transmission Lines & \checkmark & \checkmark & - & - & \checkmark & \checkmark & - & - & \checkmark & \checkmark & \checkmark & \checkmark & \checkmark & \checkmark & \checkmark & \checkmark & \checkmark & \checkmark & - & -& \checkmark   \\ 
        Solar & - & \checkmark & \checkmark & \checkmark & \checkmark & \checkmark & \checkmark & - & \checkmark & \checkmark & - & \checkmark & \checkmark & - & \checkmark & \checkmark & - & \checkmark & \checkmark & \checkmark & \checkmark   \\ 
        Wind & \checkmark & - & \checkmark & \checkmark & \checkmark & \checkmark & \checkmark & \checkmark & \checkmark & \checkmark & \checkmark & \checkmark & - & \checkmark & \checkmark & \checkmark & \checkmark & \checkmark & \checkmark & 
        \checkmark & \checkmark   \\ 
        FACTS elements & - & - & - & - & - & - & - & - & - & - & - & - & - & \checkmark & - & \checkmark & \checkmark & - & - & - & \checkmark   \\ 
        DTR & - & - & - & - & - & - & - & - & - & - & - & - & - & - & - & - & - & \checkmark & - & - & \checkmark   \\ 
        CCS retrofitting & \checkmark & - & \checkmark & \checkmark & - & \checkmark & - & - & \checkmark & - & \checkmark & \checkmark & - & - & \checkmark & - & - & \checkmark & - & - & \checkmark   \\ 
        Asset Retirement & \checkmark & - & - & - & - & - & - & - & \checkmark & - & - & - & - & - & - & - & - & - & - & - & \checkmark   \\ 
        \midrule
        \textbf{Other Considerations} & \citep{wei2022} & \citep{yin2022a} & \citep{2022} & \citep{zhao2022abc} & \citep{hou2021} & \citep{flores-quiroz2021} & \citep{lugovoy2021} & \citep{dacosta2021c} & \citep{shen2020a} & \citep{ramirez2020} & \citep{asgharian2020} & \citep{gbadamosiMultiperiodCompositeGeneration2020} & \citep{alanazi2020a} & \citep{zhou2020} & \citep{asadimajd2020} & \citep{zeinaddini-meymandDemandSideManagementBasedModel2019} & \citep{zhangCoordinationGenerationTransmission2019} & \citep{PARZEN2023121096} & \citep{lara.etal_2020} & \citep{hou.etal_2021} & This Work   \\
        \midrule
        Short-term ESS & - & - & \checkmark & \checkmark & - & \checkmark & - & \checkmark & - & - & - & - & - & - & - & - & - & - & \checkmark & \checkmark & \checkmark   \\ 
        Long-term ESS & - & - & \checkmark & \checkmark & - & \checkmark & - & \checkmark & - & - & - & - & - & - & - & - & - & - & \checkmark & \checkmark & \checkmark   \\ 
        Multi-Stage Dynamic & - & - & - & - & \checkmark & \checkmark & \checkmark & \checkmark & \checkmark & \checkmark & - & \checkmark & \checkmark & - & \checkmark & \checkmark & - & - & \checkmark & \checkmark & \checkmark   \\ 
        Geo-Spatial Input & - & - & \checkmark & \checkmark & - & - & \checkmark & - & \checkmark & - & - & - & - & - & - & - & - & \checkmark & - & - & \checkmark   \\ 
        Scalability & - & \checkmark & - & - & - & - & \checkmark & \checkmark & - & - & - & - & - & - & - & - & - & \checkmark & \checkmark & \checkmark & \checkmark   \\ 
        System sizing & - & - & \checkmark & \checkmark & \checkmark & - & \checkmark & - & \checkmark & - & \checkmark & - & \checkmark & \checkmark & \checkmark & - & - & \checkmark & \checkmark & \checkmark & \checkmark   \\  
        Annual planning stages & \checkmark & - & - & - & \checkmark & - & \checkmark & \checkmark & \checkmark & \checkmark & - & \checkmark & \checkmark & - & - & \checkmark & - & - & \checkmark & \checkmark & \checkmark   \\ 
        Parallel Technique & - & - & - & - & - & \checkmark & - & - & - & - & - & - & - & - & - & - & - & \checkmark & \checkmark & \checkmark & \checkmark   \\ 

    \bottomrule
  \end{tabular}
  }
\end{table*}

Following this guideline, in this work, we present a blueprint model using the MSP framework, capturing an unprecedented breadth of planning factors and considerations. To overcome the resulting computational complexity, the problem is decomposed using the Stochastic Dual Dynamic Programming (SDDP) algorithm \citep{pereira.pinto_1991}  leveraging high-performance computing for parallel processing. To fully capture the scope, breadth of planning factors, and considerations of this work, Table \ref{tab:LitCompare} compares this work to other similar work in the literature. Most of the long-term power system transition planning work either follows a Generation and Transmission Expansion Planning (GTEP) with primary focus on expansion of an underbuilt system or takes a reductionist approach with emphasis on a single aspect or a specific technology \citep{hou2021}. Or,  Macro Energy System (MES) planning  \citep{2022}, which often exceeds the functional boundary of the power system and does not abide by its operational constraints. For example, models like \citep{wei2022} attempt to co-optimize energy and gas systems for city-wide energy flows, leading to oversimplified networks (e.g., 24-bus and 20-node systems) and limited scenarios (6). Others like \citep{2022} \citep{HOLE2025573} completely ignore the electrical network or are geo-spatially reduced with abstract interregional flows as pseudo-branches, failing to capture physical network constraints or power flow realism \citep{lugovoy2021} \citep{lara.etal_2020} \citep{flores-quiroz2021}\citep{dacosta2021c}\citep{ramirez2020}.

On the other hand, the scope of GTEP-based work narrowness can be exemplified in work that focuses specifically on reliability metrics \citep{zhangCoordinationGenerationTransmission2019} \citep{asadimajd2020} \citep{dacosta2021c} \citep{zhou2020}, detailed unit commitment \citep{hou.etal_2021} \citep{flores-quiroz2021}, battery system modeling \citep{flores-quiroz2021}\citep{lugovoy2021}, coal retierment \citep{shen2020a} or AC flow modeling and FACTS investment \citep{zhou2020} \citep{zhangCoordinationGenerationTransmission2019}, making the transitioning problem peripheral. Achieving zero or low emissions is merely consequential in some models, such as \citep{alanazi2020a}, where the objective maximizes PV penetration with no hard carbon reduction goals. 

Another issue is in the representation of uncertainty and the stages of planning. Many models solve a deterministic problem \citep{PARZEN2023121096} \citep{lugovoy2021} or a stochastic equivalent deterministic \citep{asadimajd2020}\citep{shen2020a}\citep{gbadamosiMultiperiodCompositeGeneration2020}\citep{alanazi2020a}. Some opt for robust optimization \citep{ramirez2020} \citep{yin2022a}, sacrificing interpretability and failing to capture the adaptive value of sequential decision-making across a planning horizon. Some stochastic programming formulations are limited to two stages only \citep{zeinaddini-meymandDemandSideManagementBasedModel2019}\citep{zhou2020}. 

Many of these references are missing various essential factors, limiting the maximization of the utilization of the existing network. Instead, they increase complexity with low-value factors and little insight into infrastructural planning, such as unit commitment or deep physical modeling of a specific technology. Moreover, a lot of models allow only candidate unit/technology siting without any sizing involved, such as in \citep{zhou2020} \citep{zeinaddini-meymandDemandSideManagementBasedModel2019} \citep{zhangCoordinationGenerationTransmission2019} \citep{alanazi2020a}. Test cases where networks are modeled are often limited in size using 48-bus \citep{gbadamosiMultiperiodCompositeGeneration2020} 45-bus \citep{ramirez2020} 32-bus \citep{lugovoy2021} 24-bus \citep{zhou2020} \citep{wei2022} 20-bus \citep{asadimajd2020} 11-bus \citep{HOLE2025573} 6-bus \citep{zeinaddini-meymandDemandSideManagementBasedModel2019} \citep{hou.etal_2021} \citep{lara.etal_2020}. Some that claim large test cases (118-bus) turn out to be reducing the network to a 3-node system \citep{flores-quiroz2021} or even no network modeling at all \citep{dacosta2021c} \citep{HOLE2025573}. Models that try to capture geo-spatial planning end up reducing the renewable energy regions, limiting the potential profiles \citep{lugovoy2021}\citep{asadimajd2020} and candidate RE locations down to only three areas in some instances \citep{ramirez2020} \citep{flores-quiroz2021}, side-stepping geo-spatial correlations.


Some even focus on convexity and global optimality \citep{wei2022} \citep{HOLE2025573} \citep{hou.etal_2021}, leading these models to suffer in scalability and take a highly reductionist approach. Few adopt decomposition, and many either drop complexity or constrain scenario size \citep{wei2022}. Some work does resort to SDDP, yet the scalability of their approach is limited for various reasons, such as the focus on reliability in \citep{dacosta2021c} or \citep{lara.etal_2020} and \citep{hou.etal_2021} where the SDDiP variation is used to guarantee global optimality at the cost of increased computational burden at the subproblem level without leveraging HPC. Leveraging advanced decomposition with HPC remains confined mainly to specialized research or specific applications. A rare example where distributed computing is leveraged is in \citep{flores-quiroz2021}, which uses column generation and sharing. However, it focuses on detailed unit-commitment and storage option modeling. All the limitations are summarized in Table \ref{tab:LitCompare}.

\subsection{Contribution}

The main contribution of this work lies in introducing PSTP, a pragmatic planning framework, grounded in practice, designed for complexity, and built to scale. With some theoretical novelty manifesting in some expressions derived from incorporating FACTS, DTR, and a new transmission line into the power flow expressions and implying the Battery storage degradation/lifetime in the SDDP framework, and findings showcasing the effect of high-resolution spatio-temporal GIS-based mapping related to VRES production on the quality of the solution. The following are the features of the model presented in this work, representing a blueprint of the PSTP problem: 

\begin{itemize}
    \item \textbf{Integrating a broad spectrum of planning factors:}  relevant to short-term and long-term power system transitions. This includes the choices of Gas with Carbon Capture and Storage (GCCS) and Hydrogen Turbines (H2), Small Modular Reactors (SMR), solar and wind generation, existing unit retrofitting options, Battery Energy Storage Systems (BESS), Pumped Storage Hydroelectricity (PSH), commercially available modular solutions, such as Dynamic Thermal Rating (DTR) sensors and modular Static Synchronous Series Compensator (SSSC) devices, and finally Transmission line allocation. 
    \item \textbf{Industry-Aligned Scenario Framework with Real-World Relevance:} While stochastic programming is theoretically robust, few methodologies have used it at scale, particularly with geospatial resolution and investment-level decision granularity. This work demonstrates such a practical implementation, enabling realistic assessment of long-term uncertainty while preserving computational tractability. Furthermore, it quantifies the Value of the Stochastic Solution (VoSS) relative to standard deterministic scenario-based approaches used in the industry.
    \item \textbf{Integrated Geospatial Planning:} The model includes both wired and non-wired investment solutions and maps them spatially to the actual grid topology, offering fidelity for siting renewable generation without oversimplication of geospatial data, node aggregation, or pseudo-branch abstraction.
    \item \textbf{Scalable Decomposition Algorithm leveraging HPC and parallel processing:} This PSTP model employs a scalable decomposition algorithm based on Stochastic Dual Dynamic Programming (SDDP) with parallel computing, showcasing the viability and tractable solutions to high-dimensional, mixed-integer stochastic problems. This approach is validated on test systems representative of realistic planning complexity. 
\end{itemize}


It is important to note that this work is not merely an amalgamation of previous work. It is a methodological rethinking of transition planning and accommodating the functional boundary of power system planners aligned with the practical realities of long-term energy transition. 

The structure of this paper is as follows. Section \ref{section2} formulates and describes the PSTP model and its linearization. Section \ref{section3} describes the solution algorithm used to decompose the model and the Markov Chain representation of the scenario tree used for this model. Section \ref{section4} describes the methodologies for calculating Dynamic Thermal Rating (DTR) data, Variable Renewable Energy Resource (VRES) zone selection and outputs, scenario generation process, cost derivation, and the AESO-6 test case properties. Section \ref{section5} solves the problem six times monolithically on the AESO-6 test case, examining the effect of several planning factors on the optimal decision and the cost reduction obtained by introducing them, and evaluates the value of stochastic solution. In section \ref{section6}, the problem is solved again using the SDDP algorithm, which compares its performance to that of the monolithic solution. Then, the large test case AESO-144 is introduced and solved, demonstrating the framework’s potential scalability. Section \ref{section7} discusses the limitations of this work and planned future improvements. The paper ends with concluding remarks in Section \ref{section8}.


\section{Problem Formulation}
\label{section2}

\subsection{Model Description}

The proposed multi-stage stochastic transition planning model is formulated as a Mixed Integer Linear Problem (MILP) to optimize transition policy, technology selection, and operational schedules to meet future load forecasts, maximize existing infrastructure use, and achieve zero CO2 emissions. The model determines the optimal present capacity and technology placement while planning deferred asset allocations and operations in later stages. Asset allocation variables are referred to as transition variables, with superscripts indicating technologies: G (natural gas), N (nuclear SMR), H (hydrogen turbine), S (solar PV), W (wind), L (transmission), D (DTR sensor), F (SSSC), B (battery), P (hydro pump storage), R (CCS retrofit).

\subsection{Objective Function}

\begin{align}  
\min_{\Gamma} \ \ \ \ \ \sum_{s \in \Omega_S} \phi_s \biggl[  \sum_{y \in \Omega_Y} \biggl[ X^{inv}_{y,s} + \sum_{o \in \Omega_O} \rho_o (X^{op}_{y,s,o}) \biggr]\biggr] \tag{1} \label{1}
\end{align}

\vspace{1.5pt}

Building on existing literature, the objective function (\ref{1}) is split into two terms: the sum of investment costs $X^{inv}$ and the sum of operation costs $X^{op}$. The operational or short-term scenarios are represented in this model as discrete aggregated 24-hour scenarios multiplied by their weight $\rho_o$ at every stage. The two terms are also multiplied by the probability of long-term scenarios $\phi_s$ across the stages, representing each stage’s environmental state, such as load growth and technology costs. $\Gamma$ represents the set of all decision variables of the problem. The first term of the objective function is expanded in (\ref{2}). It includes a mix of continuous and binary variables associated with the planning factors.

\vspace{-10pt}

\begin{multline}
X^{inv} =    \sum_{l \in \Omega_{R}} (I^{L} x^{L}_l + I^{D} x^{D}_l + I^{F} x^{F}_l ) \nonumber \\
+  \sum_{n \in \Omega_B} (I^G i^G_n + I^N i^N_n + I^H i^H_n + I^P x^P_n + I^B x^B_n) \nonumber \\
+  \sum_{z \in \Omega_{Zw}} I^{W} i^W_z  +  \sum_{z \in \Omega_{Zs}} I^{S} i^S_z    
+  \sum_{g \in \Omega_G} I^R x^R_g  \tag{2} \label{2}
\end{multline} 

The investment variables in (\ref{2}) consist of installation decisions of new transmission line $x^L$ and DTR sensor $x^D$ binary variables, and the SSSC device  $x^F$ integer variable, on each Right of Way (RoW) $l$; installation decisions of candidate battery system $x^B$ and hydro-pump system $x^P$ at bus $n$; and CCS retrofitting decisions $x^R$  of existing thermal generating units $g$.
While the previous options have pre-determined capabilities, the generation technologies’ capacities are determined through the model. Namely, $i^G, i^N, i^H$ are continuous variables representing the capacity of Gas with Carbon Capture and Storage (GCSS), Nuclear SMR, and Hydrogen fuel turbines on each bus $n$, while $i^W$ and $i^S$ represent wind and solar installed capacities in each candidate VRES zone $z$. For all planning factors, $I$ is the sum of the investment cost of each technology and normalized fixed operational and maintenance costs, such as administrative fees, insurance costs, and operating labor costs in \$/p.u. capacity for the remaining years at each stage and scenario. Indeed, fixed operation and maintenance costs are included in $X^{inv}$ (in addition to investment costs), while $X^{op}$ consists of variable operation costs such as fuel and variable maintenance costs in terms of \$/p.u.

Rotary generation technologies are essential for supplying ramping, base power, grid flexibility, and inertia, as the level of VRES penetration achievable while maintaining grid reliability is not yet clear \citep{denholm2020inertia}. Thus, GCSS, SMR, and H2 are included as clean options, where GCSS is a crucial factor as it is a mature technology \citep{BUKAR2024114578}. SMRs and H2 are still developing yet promising technologies, and their availability and cost would change drastically over the years and scenarios \citep{NRELATB}, especially considering recent policies \citep{Canada_2021}; thus, including them and their respective scenarios would yield a more expansive decision space.  

Aside from these options, each planning factor has intrinsically unique properties. The retrofitting option is crucial to mitigate unnecessary new rotary unit allocation. Solar and wind energy are flagship clean energy technologies with unique environmental interaction properties. Battery and hydro-pump storage represent short-term (or storage as transmission) and long-term energy capacity solutions, respectively \citep{rawa2022}, with hydro-pump storage interacting with the environment, leading to the stochastic behavior of the reservoir level. DTR sensors provide real-time thermal capacity of the transmission lines depending on the weather conditions, and their monitored data can affect utilization rates, VRES allocation, and congestion mitigation strategies. Similarly, the SSSC device provides a modular solution for actively managing the network flow by creating virtual reactance on the lines. Finally, transmission line allocation is added as a last resort, as it might be the only viable solution for specific networks.

The objective function's operational cost term $X^{op}$ expanded in (\ref{3}) involves hourly operation outputs and costs. It includes constants $P$, the operation costs in \$/p.u., CO2 tax costs $P^{E_{CO2}}$ and $P^{R_{CO2}}$ for thermal units before and after retrofitting, respectively, are based on estimated emissions p.u. of production. The outputs of the existing thermal units are denoted by  $p^{E}_{g}$ and $p^{C}_{g}$ before and after CCS retrofitting, respectively, for every unit $g$. Depending on the retrofitting state, only one of these two is active. $p^G_n$, $p^N_n$, and $p^H_n$ are the generation outputs of GCSS, SMR, and H2 at each bus $n$. The magnitude of load-shedding at each bus $n$ is represented by $p^L_n$. Wind and solar curtailment values at each zone $z$ are denoted by $p^W_z$ and $p^S_z$, respectively. Battery degradation is modeled and considered operationally, preventing the need to include battery degradation costs. 

\vspace{-10pt}

\begin{multline}
    X^{op} = 
    \sum_{t \in \Omega_{T}} \bigg( \sum_{g \in \Omega_{G}} \Big[(P^{E} +P^{E_{CO2}})p^{E}_{g} 
+  (P^{R} +P^{R_{CO2}})p^{C}_{g}\Big] \nonumber \\
+  \sum_{n \in \Omega_{B}} \big( P^{G} p^{G}_{n} +  P^{N} p^{N}_n + P^{H} p^{H}_n + P^{L}  p^{L}_{n}  \big) \nonumber \\
+   \sum_{z \in \Omega_{Zw}} P^{W}p^{W}_{z} + \sum_{z \in \Omega_{Zw}} P^{S} p^{S}_{z}\bigg)
\tag{3} \label{3}
\end{multline}

\vspace{-10pt}

\subsection{Transition Decision Constraints}

These constraints are imposed on the transition variables of the problem. \\

\vspace{1.5pt}

\subsubsection{Maximum New Connection Capacity of Substations}

\begin{multline}
 \sum_{\tau \le y} (i_{n,\tau,s}^G + i_{n,\tau,s}^N + i_{n,\tau,s}^H) + \sum_{\tau \le y}  \sum_{z \in \Omega_{Zs,n}} i_{z,\tau,s}^S \nonumber \\ 
+  \sum_{\tau \le y}  \sum_{z \in \Omega_{Zw,n}} i_{z,\tau,s}^W \le G_{n,y}^{max} \ \ \ \forall{n,y,s} \tag{4} \label{4}
\end{multline}

\vspace{1.5pt}

This constraint limits the new p.u. capacity connected to each bus $n$. It means that, at every stage, the total installed capacity cannot exceed the p.u. limit of $G_n^{max}$. This includes the sum of VRES across zones $z$ associated to a particular bus $n$.\\

\subsubsection{Lifetime}

\begin{align}
  1 & \ge \sum_{\tau = y}^{y + \gamma} x_{l,\tau,s}^{\{D,F\}} \ \ \ \forall{l,s} \ for \ y = 1,...,Y - \gamma  \tag{5a} \label{5a} \\
  1 & \ge \sum_{\tau = y} ^ Y x^{\{D,F\}}_{l,\tau,s} \ \ \ \forall{l,s}  \ for \  y = Y-\gamma+1,...,Y\ \tag{5b} \label{5b}\\
  1 & \ge \sum_{\tau = y}^{y + \gamma} x^{\{B,P\}}_{n,\tau,s} \ \ \ \forall{n,s} \ for \ y = 1,...,Y - \gamma  \tag{6a} \label{6a} \\
   1 & \ge \sum_{\tau = y} ^ Y x^{\{B,P\}}_{n,\tau,s} \ \ \ \forall{n,s}  \ for \  y = Y-\gamma+1,...,Y\ \tag{6b} \label{6b}\\
  1 & \ge \sum_{\tau = y}^{y + \gamma} x^R_{g,\tau,s} \ \ \ \forall{g,s} \ for \ y = 1,...,Y - \gamma  \tag{7a} \label{7a} \\
   1 & \ge \sum_{\tau = y} ^ Y x^R_{g,\tau,s} \ \ \ \forall{l,s}  \ for \  y = Y-\gamma+1,...,Y\ \tag{7b} \label{7b}
\end{align}

\vspace{1.5pt}

In equations (\ref{6a} - \ref{7b}), $Y$ is the last stage of the planning horizon $\max(\Omega_Y)$. For technologies where transition variables represent a set of prospective future options, it is imperative to ensure that each asset incurs a cost on the objective function solely at the time of allocation while its operational impact is sustained throughout its entire lifespan. Most jurisdictions aim to decarbonize the grid by 2035-2050 \citep{IEA2021}. This implies that, for any given future date, the lifespan of planning factors (except battery systems) may exceed the maximum planning horizon considered in this model ($\gamma \ge Y$) \citep{en16114516, JORDAAN2021100058, Jorge2011LifeCA}. In that case, subequations ``a" can be omitted.\\

\subsubsection{Line-Dependant Allocation}

\begin{align}
x^m_{l,y,s} \le \sum_{\tau \le y}  x^L_{l,\tau,s} + N_l \ \ \ \forall{l,y,s} \ \  \forall{m} \in \{D, F\}  \tag{8} \label{8} 
\end{align}

\vspace{1.5pt}

The constraints (\ref{8}) ensure that the DTR sensor and SSSC device allocation can only be done on a RoW $l$ if it has an existing or invested line. Superscript $m$ represents the specific technology with the variable  $x$ (e.g., $D$ for DTR and $F$ for FACTS). This notation is employed in the paper for brevity and readability wherever applicable. \\

\subsubsection{Available Area for New Technologies}

\begin{align}
& \sum_{\tau \le y}  K^mi^m_{n,\tau,s} \le A_{n,y}^m \ \ \ \forall{n,s,y} \ \ \forall{m} \in \{G,H,N\} \tag{9} \label{9} \\ 
& \sum_{\tau \le y}  K^mi^m_{z,\tau,s} \le A_{z,y}^m \ \ \ \forall{z,s,y} \ \ \forall{m} \in \{S,W\} \tag{10} \label{10}  
\end{align}

\vspace{1.5pt}

Constraint (\ref{9}) is for thermal generation technologies, and (\ref{10}) is for VRES technologies. These constraints limit the capacity invested by the available area $A$ at bus $n$ or zone $z$. $K$ is the area occupied per p.u. capacity for each technology. This constraint is especially crucial for VRES, as each zone has a different output capacity within a limited area. 

\subsection{Operational Constraints}

\vspace{1.5pt}

Previous-stage transition variables limit the current stage operations, i.e., new technology allocation decisions from previous stages take effect at the operation of the current stage. Thus, operational constraints for the new equipment start taking effect from $y=2 ...Y$, where $Y$ is the last stage of the planning horizon. The operational and random variables in all the expressions can be distinguished from transition variables by their subscript $t$. 

\vspace{7pt}

\subsubsection{ Maximum Thermal Power Outputs}

\begin{align}
\Big\{ & P^{Min}(1 - \sum_{\tau < y}x_{g,\tau,s}^R) \le p_{g,t,o,y,s}^E \ \forall{g}  \tag{11a} \label{11a}\\
& P^{Max}(1 - \sum_{\tau < y}x_{g,\tau,s}^R) \ge p_{g,t,o,y,s}^E  \ \forall{g} \tag{11b} \label{11b}\\
& P^{Min}\sum_{\tau < y}x_{g,\tau,s}^R  \le p_{g,t,o,y,s}^R \le P^{Max} \sum_{\tau < y}x_{g,\tau,s}^R   \ \forall{g}
  \tag{12} \label{12}\\
& G^{Min} \sum_{\tau < y} i^m_{n,\tau,s}  \le p^m_{n,t,o,y,s} \le G^{Max} \sum_{\tau < y} i^m_{n,\tau,s} \ \forall{n}\tag{13} \label{13} \\ 
& \ \ \forall{m} \in \{G,H,N\} \Big\} \ \ \ \ \forall{t,o,y,s}   \nonumber
\end{align}

The existing units' maximum and minimum outputs in (\ref{11a}), (\ref{11b}), and (\ref{12}) are imposed by their given parameters. The output limits of new rotary generation in  (\ref{13}) are defined by the amount of capacity invested in previous stages up to the current stage, multiplied by production factors $ G^{Max}\ \& \ G^{Min}$ that are typical of their respective technologies \citep{nrel2020,osti_2371533}. In (\ref{11a}), (\ref{11b}) and (\ref{12}), if existing units are retrofitted (i.e., $x^R$ becomes active), then the $p^R$ is active, and $p^E$ is forced to zero and vice versa. That way, different power output emission levels and operational costs can be assigned upon retrofitting. \\

\vspace{1.5pt}

\subsubsection{Target Carbon Emission}

\begin{multline}
\sum_{t \in \Omega_t} \sum_{g \in \Omega_G} (CO2^E_{g}  \ (p^E_{g,t,o,y,s}) \nonumber \\  +  \  CO2^R_{g} \  (p^R_{g,t,o,y,s})) \le T_y \ \ \ \forall{o,y,s} \tag{14} \label{14} 
\end{multline}

This constraint is key to the transition plan. The output of CO2-emitting generators has to be limited to a specific level $T$ in terms of tonnes of CO2 emissions at every stage $y$. $CO2^E_{g}$ and $CO2^R_{g}$ are the tonnes/p.u. emissions before and after retrofitting, respectively

\vspace{7pt}

\subsubsection{Load Shedding} 
\begin{align}
p^L_{n,t,o,y,s} \le \zeta^D_{n,t,o,y,s}  \ \ \ \forall{n,t,o,y,s} \tag{15} \label{15}
\end{align}

The amount of load shedding cannot exceed the demand realization $\zeta^D$. The planners can control the degree of load shedding by simply turning the values of $\zeta^D$ to a constant or zero if no load shedding is desired.\\

\vspace{1.5pt}

\subsubsection{RE Curtailment}

\begin{multline}
 p^m_{z,t,o,y,s} \le \zeta^m_{z,t,o,y,s}(R^m_z + \sum_{\tau < y} i^m_{z,\tau,s}) \\  \forall{z,t,o,y,s} \ \   \forall{m} \in \{S,W\} \tag{16} \label{16}  
\end{multline}

The amount of VRES curtailment cannot exceed the available VRES output, which is the solar and wind production factor realization ($\zeta^S$ and $\zeta^W$) multiplied by the summation of the existing VRES capacity and installed VRES capacity in the previous stages. $\zeta^S$ and $\zeta^W$ are randomly determined for each hour, short and long-term scenarios, and VRES zone.  \\

\vspace{1.5pt}

\subsubsection{Ramping Up and Down}

\begin{align}
\Big\{ & DN_g \le p^E_{g,t,o,y,s} - p^E_{g,t-1,o,y,s} \le UP_g   & \forall{g} \tag{17} \label{17} \\
& DN_g \le p^R_{g,t,o,y,s} - p^R_{g,t-1,o,y,s} \le UP_g \ & \forall{g} \tag{18} \label{18} \\
& DF^m \sum_{\tau < y} i^m_{n,\tau,s} \le p^m_{n,t,o,y,s} - p^m_{n,t-1,o,y,s} & \forall{n} \tag{19a} \label{19a} \\
& UF^m \sum_{\tau < y} i^m_{n,\tau,s} \ge p^m_{n,t,o,y,s} - p^m_{n,t-1,o,y,s} & \forall{n} \tag{19b} \label{19b} \\ & \ \ \ \forall{m} \in \{G,H,N\} \Big\}  \ \ \ \forall{o,y,s}, \  \forall t > 1 \nonumber
\end{align}

Ramping limits of existing units before and after retrofitting (\ref{17}) and (\ref{18}) are determined by their given parameters ($UP, DN$). Limits of new rotary units (\ref{19a}) and (\ref{19b}) are determined by the invested capacity multiplied by factors ($UF, DF)$ specific to the technology.    \\

\subsubsection{Nodal Balance}

\begin{multline}
   \sum_{l|j(l)=n} f_{l,t,o,y,s,j}  - \sum_{l|k(l)=n} f_{l,t,o,y,s,k} \nonumber \\ 
+  \sum_{z \in \Omega_{Zw,n}} (\zeta^W_{z,t,o,y,s}[R_{z}^{W} + \sum_{\tau < y}i_{z,\tau,s}^{W}] - p_{z,t,o,y,s}^{W}) \nonumber  \\
+  \sum_{z \in \Omega_{Zs,n}} (\zeta^S_{z,t,o,y,s}[R_{z}^{S} + \sum_{\tau < y}i_{z,\tau,s}^{S}] - p_{z,t,o,y,s}^{S}) \nonumber \\
+  \sum_{g \in \Omega_{Gn}} (p_{g,t,o,y,s}^{E} + p_{g,t,o,y,s}^{C}) + p_{n,t,o,y,s}^{G} + p^{N}_{n,t,o,y,s}  \nonumber \\
+  p^{H}_{n,t,o,y,s}  + p_{n,t,o,y,s}^{DI} - p_{n,t,o,y,s}^{CH} + p_{n,t,o,y,s}^{T}  \nonumber \\  
-  p_{n,t,o,y,s}^{P} + \ p_{n,t,o,y,s}^{L} - \zeta^D_{n,t,o,y,s} = 0 \tag{20} \label{20}
\ \ \ \forall n,t,o,y,s \end{multline}  

\vspace{1.5pt}

The nodal balance constraint imposes the active power balance at each system node. In the first two terms on the left side of (\ref{20}), $f_j$ is the incoming flow to bus $n$, and $f_k$ is the outgoing flow from bus $n$. In this model, network flows are modeled using the DC power flow formulation \citep{DCPFlow}, which is widely used and considered an appropriate approximation in GTEP and MES planning \citep{PARZEN2023121096}. The third and fourth terms are related to the power injection to bus $n$ from VRES. Allocations of  VRES technology in the proposed model occur in geographical zones. Each zone is associated with a bus through the sets $\Omega_{Zs,n}$ and $\Omega_{Zw,n}$. As implied by the third and fourth terms, the available VRES output minus the VRES curtailment in every node indicates the VRES injection. The fourth line in (\ref{20}) represents the power output of existing units before and after retrofitting and new units at bus $n$. $p^{DI}$ and $p^{CH}$ are the battery discharging and charging power at bus $n$. $p^{T}$ and $ p^{P}$ are the hydropump turbining and pumping energy at bus $n$. The last line shows the load-shedding amount and the demand random variable $\zeta^D$.  \\

\subsubsection{ Bus Angle Limit }
\begin{align}
     \theta^{min} \le \theta_{n,t,o,y,s} \le \theta^{max} \ \ \ \forall{n,t,o,y,s} \tag{21} \label{21}
\end{align}

Typical DC bus voltage limits impose bounds on bus voltage angles; this limit depends on many factors and can be system specific but generally would not exceed $\theta = \pm 30$° \citep{glover2012power}. \\

\subsubsection{Power Flow Rule}
\begin{align} 
f_{l,t,o,y,s} = \frac{1}{X_l} (\theta_{j(l)} - \theta_{k(l)}) +  \Delta f_{l,t,o,y,s} \ \ \ \forall{l,t,o,y,s}  \tag{22} \label{22}
\end{align}

The main difference of (\ref{22}) with DC power flow formulation is the addition of the SSSC device injection variable $\Delta f$. If no SSSC is allocated, this variable takes the value of 0. The constraints defining $\Delta f$ boundaries are addressed later. \\

\subsubsection{Battery Storage}

\begin{align}
\Big\{  \ \ 0 & \le p^{CH}_{n,t,o,y,s} \le CH^{max}x_{n,y-1,s}^{B}  \tag{23} \label{23}\\
 0 & \le p^{DI}_{n,t,o,y,s} \le DI^{max}x_{n,y-1,s}^{B} \tag{24} \label{24} \\
 0 & \le p^{CH}_{n,t,o,y,s} \le CH^{max}(x_{n,t,o,y,s}^{state})  \tag{25} \label{25}\\
 0 & \le p^{DI}_{n,t,o,y,s} \le DI^{max}(1 - x_{n,t,o,y,s}^{state}) \tag{26} \label{26} \\
 SOC^{min} & \le s_{n,t,o,y,s} \le  SOC^{max} \tag{27} \label{27} 
\ \ \Big\} \ \ \forall{n,t,o,y,s}  
\end{align}

\vspace{-23pt}

\begin{multline}
 s_{n,t+1,o,y,s}  = s_{n,t,o,y,s} + \eta^{CH} p^{CH}_{n,t,o,y,s} \\ 
 - \eta^{DI} p^{DI}_{n,t,o,y,s} \tag{28}  \label{28} \ \ \forall{n,o,y,s} \ \forall t < \text{max}(\Omega_T) 
\end{multline}

Constraints (\ref{23}) and (\ref{24}) make the charging and discharging rates of the battery system zero if no allocation is made (if $x^{B}$ is 0) and limit charging and discharging rates to their maximum limits if the allocation is made (if $x^{B}$ is 1). Constraints (\ref{25}) and (\ref{26}) avoid simultaneous charging and discharging of the battery system using the binary variable $x^{state}$. Constraint (\ref{27}) limits the State of Charge (SOC), and (\ref{28}) presents the dynamic power balance constraint of the battery. It is important to note that since the operation is modeled hourly, there is no difference between power and energy in the storage models. \\

\subsubsection{Battery Degredation} 

\begin{align}
\Big\{ & D_{n,t,o,y,s}^{cy} \ge -0.00102 \ \frac{s_{n,t,o,y,s}}{SOC^{max}} + 0.00051 \ \  \forall{t} \tag{29a} \label{29a}\\
& D_{n,t,o,y,s}^{cy} \ge -0.000151 \ \frac{s_{n,t,o,y,s}}{SOC^{max}} + 0.00015  \ \ \forall{t} \tag{29b} \label{29b} \\ 
& \sum_{t} \big( D_{n,t,o,y,s}^{cy} + D_{t}^{shelf} \big) \le \frac{1 - \epsilon^B} {\gamma} \ \Big\} \ \ \ \forall{n,o,y,s} \tag{30} \label{30}
\end{align}

\begin{figure}[ht]
    \centering
    \includegraphics[width=0.70\textwidth]{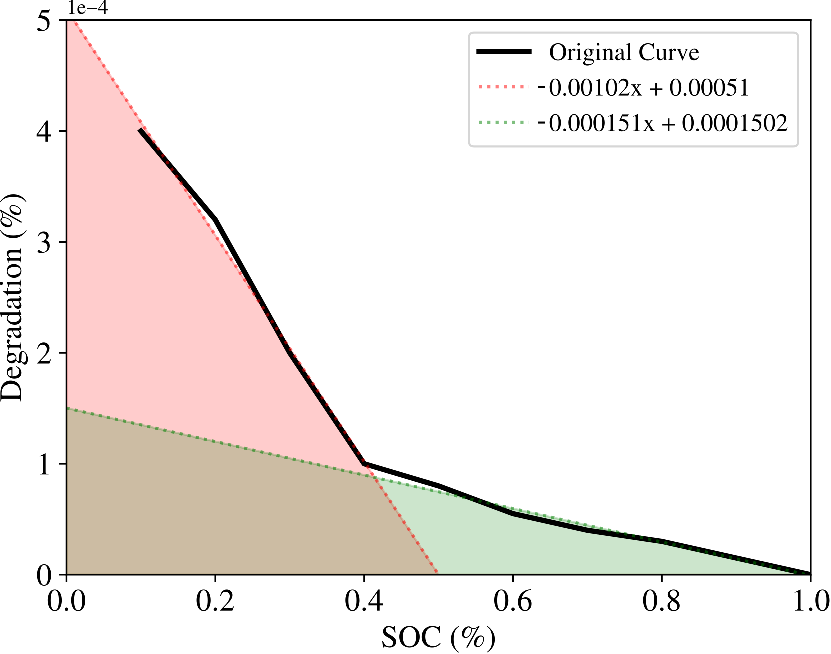}
    \caption{Degradation vs. SOC curve and its linear approximation}
   \label{fig:battery}
\end{figure}

To model battery degradation based on the depth of discharge, the degradation versus SOC curve is used \citep{8946818}. The curve is linearly fitted to $R^2 >0.98$ using two equations as shown in Fig. \ref{fig:battery}. The degradation rate can then be limited using the lines in Fig. \ref{fig:battery}, based on the desired battery lifetime $\gamma$, by the system planner as shown in (\ref{29a}-\ref{30}). 

$\epsilon^B$ is the percentage criteria of the maximum state of charge $SOC^{max}$ that indicates the battery’s End-of-life. $D^{cy}$ is the hourly operational degradation and $D^{shelf}$ is the shelf degradation of the battery system \citep{8946818}. This way, battery operation throughout stages is limited to guarantee the desired lifetime. This approach offers a novel framework for modeling battery degradation, grounded in the practical realities system planners face. \\

\subsubsection{Hydropump Storage}

\begin{align}
\Big\{ \ \ \ \  p^T_{n,t,o,y,s} & = \sigma^T w_{n,t,o,y,s}^T \tag{31} \label{31} \\
 p^P_{n,t,o,y,s} & = \sigma^P w_{n,t,o,y,s}^P \tag{32} \label{32} \\
 v^U_{n,t+1,o,y,s} & = v^U_{n,t,o,y,s} + w^P_{n,t,o,y,s} - w^T_{n,t,o,y,s} \tag{33} \label{33} \\
 v^L_{n,t+1,o,y,s} & = v^L_{n,t,o,y,s} + w^T_{n,t,o,y,s} - w^P_{n,t,o,y,s} \tag{34} \label{34} \\
 V^{U,min} & \le v_{n,t,o,y,s}^U \le V^{U,max} \tag{35} \label{35} \\ 
 V^{L,min} & \le v_{n,y,o,y,s}^L \le V^{L,max} \tag{36} \label{36} \\ 
 V^{U,0} & \le  v^U_{n,T} \ \   \tag{37} \label{37} \\ 
 V^{L,0} & \le v^L_{n,T}  \ \  \tag{38} \label{38} \\ 
 0 & \le w_{n,t,o,y,s}^T \le W^{max}\sum_{\tau < y}  x_{n,\tau,s}^{P} \tag{39} \label{39}\\
 0 & \le w_{n,t,o,y,s}^P \le W^{max}\sum_{\tau < y}  x_{n,\tau,s}^{P} \tag{40} \label{40}\\
& \Big\}\ \ \ \ \forall{n,t,o,y,s} \nonumber
\end{align}

The main distinction between this model and the battery model is in the formulation of the water flow and reservoirs, which, unlike the battery system, makes it subject to climate stochasticity (i.e., the water flow can be treated as a random variable). Constraints (\ref{31}) and (\ref{32}) define the power produced/consumed as a function of the turbine/pump water flow with a conversion factor $\sigma$. Constraints (\ref{33}) and (\ref{34}) determine the water volume of the upper and lower reservoirs in the next period (water balance constraints). Constraints (\ref{35}) and (\ref{36})  determine the maximum and minimum limits for the upper and lower reservoirs. Constraints (\ref{37}) and (\ref{38}) set the end coupling constraints for both reservoirs to avoid depleting them in a set $T$ of operation periods. Finally, constraints (\ref{39}) and (\ref{40}) limit the turbine and pump water flows. When no allocation is made ($x^P$ is 0 in the previous stages), the turbine and pump water flows become zero. \\

\vspace{1.5pt}

\subsubsection{Non-Anticipativity Constraints}

\begin{align}
\Big\{& i^m_{n,y,s} = i^m_{n,y,\tilde{s}}  & \forall{n}  \ \forall{m} \in \{G,N,H\}   \tag{41} \label{41} \\
& i^m_{z,y,s} = i^m_{z,y,\tilde{s}}  & \forall{z}  \ \forall{m} \in \{S,W\} \tag{42} \label{42} \\
& x^m_{n,y,s} = x^m_{n,y,\tilde{s}}  & \forall{n}  \ \forall{m} \in \{B,P\} \tag{43} \label{43} \\
& x^m_{l,y,s} = x^m_{l,y,\tilde{s}}  &  \forall{l} \ \ \forall{m} \in \{L,D,F\} \tag{44} \label{44} \\
& x^R_{g,y,s} = x^R_{g,y,\tilde{s}}  & \forall{g} \tag{45}  \label{45} \Big\} \ \ \forall{y,s,\tilde{s}} \  \| \zeta_{y,s} = \zeta_{y,\tilde{s}}
\end{align}

These constraints preserve the tree structure of the multistage stochastic model \citep{Higle_2005}. If two different child nodes of the scenario tree come from the same parent node, then the transition decision of their parent node (previous stage decision) has to be the same. Expression (\ref{41}) is for gas, H2, and SMR transition variables. Expression (\ref{42}) is for solar and wind transition variables. Expression (\ref{43}) is for storage transition variables. Expression (\ref{44}) is for transmission reinforcement transition variables. Expression (\ref{45}) is for the retrofitting option.    \\

\subsubsection{Transmission Line Thermal Rating}

\begin{align}
f_{l,t,o,y,s}  -  S^{ST,E}_l & \le 0 \nonumber  \ \ \ \ \forall{l,t,o,y,s} \tag{46a} \label{46a} \\ 
- f_{l,t,o,y,s} -  S^{ST,E}_l & \le 0 \nonumber \ \ \ \ \forall{l,t,o,y,s}  \tag{46b} \label{46b} 
\end{align} 

The typical line thermal rating constraint only involves the static line thermal rating and the line flow variable, as shown in (\ref{46a}) and (\ref{46b}). New lines in this model would have pre-determined properties. Taking (\ref{46a}) for an example, the expression changes to:

\begin{align}
f_{l,t}  - & S^{ST,N}_l\sum_{\tau < y} x^L_l + S^{ST,E} \le 0 \nonumber \ \ \ \ \forall{l,t,o,y,s} \tag{47} \label{47} 
\end{align} 

Adding the DTR planning factor means line capacities switch from the static rating to the DTR random variable $S^{DTR}$ when $x^D$ is active. Then (\ref{46a}) transforms to (similarly for (\ref{46b})): 

\begin{multline}
f_{l,t,o,y,s}  -  \big(S^{ST,N}_l\sum_{\tau < y} x^L_{l,\tau,s} + S^{ST,E}\big)(1 - \sum_{\tau < y} x^D_{l,\tau,s}) \nonumber \\
-  \big(S^{DTR,N}_l\sum_{\tau < y} x^L_{l,\tau,s} + S^{ST,E}\big)\big(\sum_{\tau < y} x^D_{l,\tau,s}\big) \le 0  \nonumber \ \ \ \\  \forall{l,t,o,y,s} \tag{48} \label{48}     
\end{multline}

\vspace{1.5pt}

This is non-linear as it contains binary multiplications in the form of $\sum_{\tau \le y} x^L_{l,\tau,s}\sum_{\tau \le y} x^D_{l,\tau,s}$. It can be linearized by introducing a new binary variable $v_{l} \in \{0,1\}$ for every stage and scenario. With this replacement and the addition of a few constraints on the added variable, (\ref{49}-\ref{53}) are obtained as the linear equivalent of (\ref{48}):

\begin{multline}
f_l -  S^{ST,N}_l\big(\sum_{\tau \le y} x^L_{l,\tau,s} - \sum_{\tau \le y} v_{l,\tau,s}\big) + S^{ST,E}\big(1 -\sum_{\tau \le y} x^D_{l,\tau,s}\big) \nonumber \\ 
-  \big(S^{DTR,N}_l\sum_{\tau \le y} v_{l,y,s} + S^{DTR,E}\sum_{\tau \le y} x^D_{l,y,s}\big) \le 0  \ \  \forall{l,t,o,y,s} \tag{49}  \label{49} 
\end{multline}

\vspace{-15pt}

\begin{align}
\Big\{ \ \   v_{l,y,s} & \le x_{l,y,s}^L \ \ \  \tag{50} \label{50} \\
   v_{l,y,s} & \le x_{l,y,s}^D \ \ \ \tag{51} \label{51}\\
   v_{l,y,s} & \ge x_{l,y,s}^L + x_{l,y,s}^D - 1 \ \  \tag{52} \label{52} \\
   v_{l,y,s} & \in \{0,1\} \quad \quad \Big\}  \quad  \forall{l,y,s}  \tag{53} \label{53} 
\end{align}

This represents a novel modeling approach to transmission line capacity that accounts for the interplay of multiple planning factors and the flexibility in allocation decisions.

\subsubsection{ SSSC Injection factor and linearization} 

\begin{align}
\Delta f_{l} = \frac{1}{\Delta X_{l}}(\theta_{j(l)} - \theta_{k(l)}) \tag{54} \label{54}
\end{align}

The function (\ref{54}) defines the injection factor variable $\Delta f_l$. The SSSC device does not change reactance directly but injects voltage to emulate reactance change $\Delta X_l$, changing the flow on the line. Given an SSSC device voltage range, the emulated reactance depends on the current magnitude flowing through the line. Fig. \ref{fig:Area} shows the operational properties of a sample SSSC device. For a given device, a series of reformulations in \citep{Rui_Nudell_2021} show that $\Delta f$ limits can be defined as: 

\begin{align}
-N^F_l V |\frac{1}{X_l}| \le \Delta f_l \le N^F_l V |\frac{1}{X_l}| \tag{55} \label{55} \ \ \ \ 
\end{align}

Where $N^F_l $ is the number of modular SSSC devices to be installed on the line $l$, and V is the p.u. voltage capability of the SSSC device \citep{Rui_Nudell_2021}. $N^F_l $ is replaced by the variable $x^F_l$, a positive integer variable in this case. 

\begin{align}
- x^F_l V |\frac{1}{X_l}| \le \Delta f_{l,t} \le x^F_lV |\frac{1}{X_l}|\ \ \  \forall{l,t} \ \  x^F_l \in \mathbb{Z^+}  \tag{56} \label{56}
\end{align}

Also, the device has a current magnitude cut-in limit of $C$, where it can only operate if it is reached. To model this condition, first, it is recalled that in DC power flow $|I| \approx |f|$. Then, the cut-in conditions can be stated as: 

\vspace{-10pt}

\begin{align}
 & \Delta f_{l,t} = 0 \ \ \ \ \ | \  f \le C \tag{57} \label{57} \\ 
 & \Delta f_{l,t} = 0 \ \ \ \ \ |  \ f \ge -C  \tag{58} \label{58}
\end{align}

\begin{figure}[t!]
    \centering
    \includegraphics[width=0.90\textwidth]{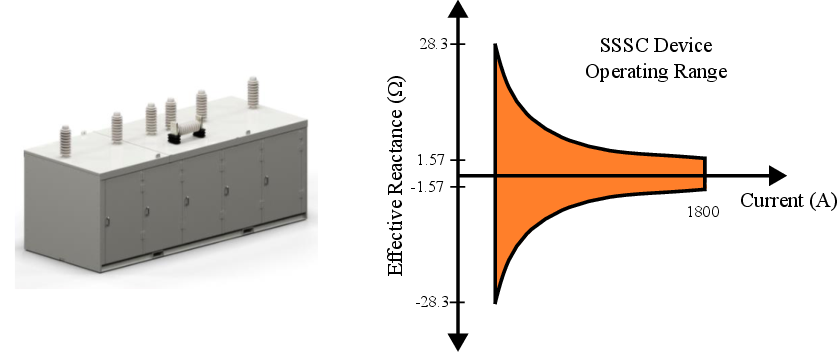}
    \caption{The SSSC device characteristics, field-proven by SmartWire$^{TM}$ \citep{SmartWiresInc._2023}}
   \label{fig:sssc}
\end{figure}

The cut-in conditions can be modeled with the addition of two binary variables $u^a$ and $u^b$: 

\begin{align}
\Big\{ \ \ & -x^F_l V |\frac{1}{X_l}| \le  \Delta f_l \le x^F_l V |\frac{1}{X_l}| \tag{59} \label{59}  \\
 - & M^{\Delta}|\frac{1}{X_l}|(u_{l,t,o,y,s}^a + u_{l,t,o,y,s}^b)\le  \Delta f_{l,t,o,y,s} \tag{60a} \label{60a}\\
& M^{\Delta}|\frac{1}{X_l}| (u_{l,t,o,y,s}^a + u_{l,t,o,y,s}^b)  \ge \Delta f_{l,t,o,y,s} \tag{60b} \label{60b} \\
& f_{l,t,o,y,s} - C \le M^{f}u_{l,t,o,y,s}^a  \tag{61} \label{61}  \\
& C - f_{l,t,o,y,s}  \le M^{f}(1-u_{l,t,o,y,s}^a)  \tag{62} \label{62} \\
- & C -f_{l,t,o,y,s} \le M^{f}u_{l,t,o,y,s}^b   \tag{63} \label{63} \\
& f_{l,t,o,y,s} + C \le M^{f}(1-u_{l,t,o,y,s}^b) \Big\}  \quad \forall{l,t,o,y,s} \tag{64} \label{64} 
\end{align}

The big $M^{\Delta}$ should be set to a value larger than the upper bound of $x^F_l$, whereas big $M^{f}$ can be set as the $M^{f}=2 \times \max(S^{DTR}, S^{ST})$. Expressions (\ref{61}) and (\ref{62}) ensure the activation of $u^a$ if $f \ge C$ and (\ref{63}) and (\ref{64}) activate $u^b$ if $f \le -C$. If neither variable is activated, $\Delta f$ will be 0 according to (\ref{60a}) and (\ref{60b}).  To our knowledge, this is the first planning framework to model these operationally essential cut-in conditions explicitly.

\section{SDDP Algorithm and Markovian Representation}
\label{section3}

 Pereira originated the Stochastic Dual Dynamic Programming (SDDP) algorithm to solve hydrothermal operation planning problems \citep{pereira.pinto_1991}. As a Benders-type method, SDDP constructs linear approximations of cost-to-go functions to solve multistage stochastic problems efficiently. Unlike Nested Decomposition \citep{Vanderbeck_2001}, SDDP is sampling-based, which avoids the exponential growth of scenarios in a decision tree and enhances scalability. It has since been adapted to diverse domains, including multi-market scheduling \citep{helseth.etal_2016}, gas transportation \citep{toledo.etal_2016}, unit commitment \citep{alvarez.etal_2019}, and GTEP \citep{hou.etal_2021}. Numerous extensions exist, including risk-averse \citep{liu.shapiro_2020}, robust and distributionally robust formulations \citep{Matos2018, morillo.etal_2022}, and mixed-integer adaptations with relaxed integer constraints \citep{zhou2020}.

Despite its power, SDDP can still become computationally expensive with large scenario trees, particularly when higher-order autoregressive processes represent time-dependent uncertainty. To mitigate this, we adopt a Markov Chain-based variation known as MC-SDDP \citep{LOHNDORF2019650}. In this approach, operational decisions are embedded within intrastage subproblems, where a node at a given stage can transition to any node in the next, enabling a dramatic reduction in the number of nodes while using a single expected cost-to-go function per stage instead of one per node \citep{dowson2021sddp}. 
The Markovian structure requires that the data process obey a memoryless property, making it suitable only for problems with collapsible scenario trees where all paths can lead to equivalent states.
Short-term operational scenarios that are not sequentially linked are modeled using 24-hour representative days of a year operation to ensure stage-wise independence. These representative days result from a first-order, periodic autoregressive exogenous Markov process, which decouples scenarios across stages.  The differences between short-term scenarios across nodes are modeled by applying state-dependent scaling coefficients \citep{GUEVARA2024110084}, derived from long-term scenario outlooks such as NREL’s ATB \citep{NRELATB} and the Shared Socioeconomic Pathways (SSP) \citep{ipcc2021ar6}.

\begin{figure}[ht] 
\centering \includegraphics[width=0.70\textwidth]{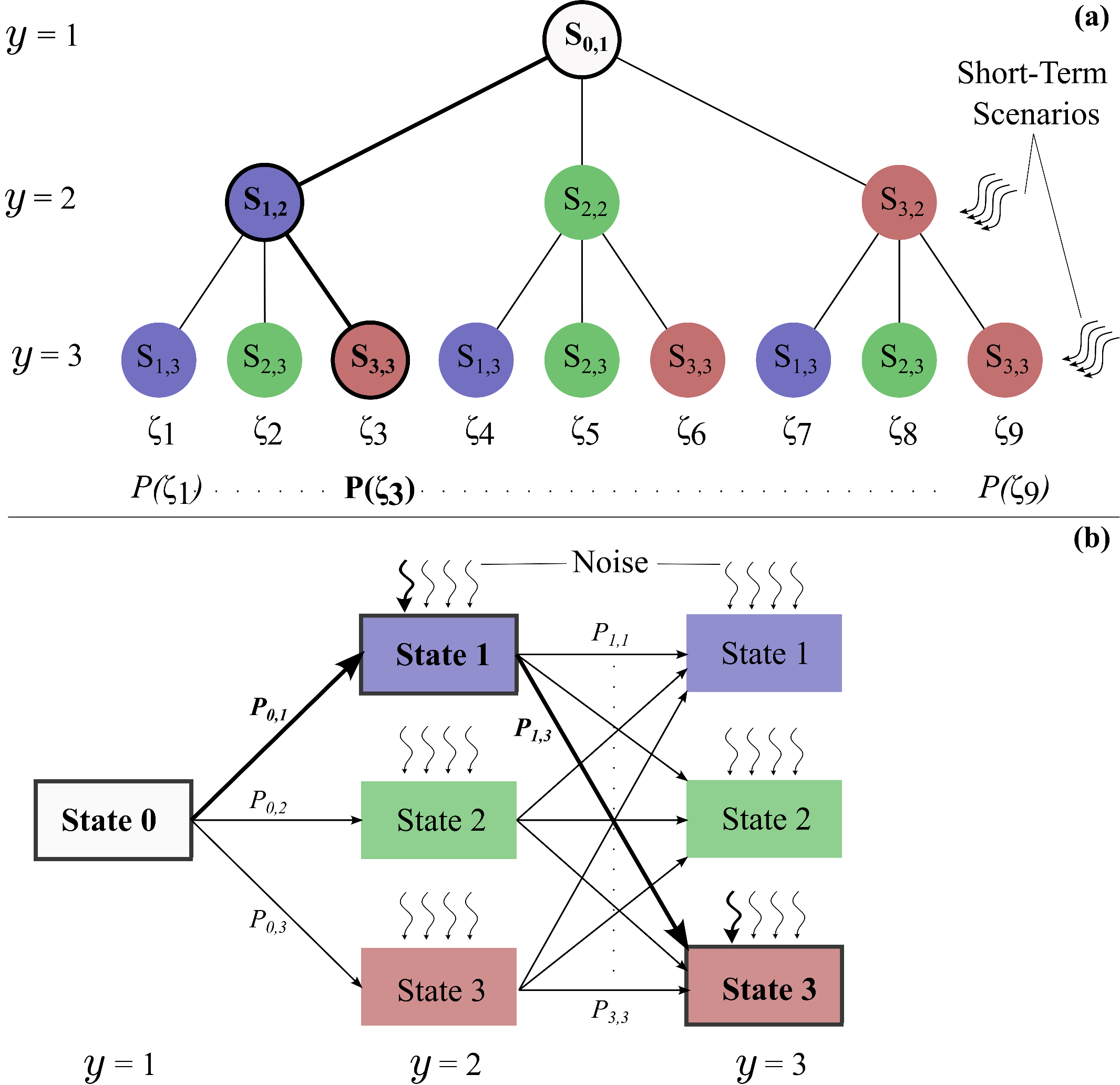} \caption{(a) Scenario tree representation. (b) Markov chain representation.} \label{fig:MCchain} \end{figure}

Figure \ref{fig:MCchain} illustrates how the scenario tree collapses into a Markov Chain representation. In the tree (Fig. \ref{fig:MCchain}a), each node $S_{s,y}$ reflects the combination of short-term and long-term uncertainties at stage $y$. Nodes within a stage with the same short-term characteristics are collapsed (Fig. \ref{fig:MCchain}b), and transitions between states follow Markovian probabilities. In this formulation, a scenario becomes a sequence of nodes and ``noise" (short-term) realizations, where each node connects to all nodes in the next stage.

This is a commonly accepted assumption in multistage planning literature \citep{lara.etal_2020}, and while it precludes modeling time-dependent uncertainty with full fidelity, it offers significant computational advantages. Moreover, interstage dependencies for long-term processes, such as annual peak load uncertainty, can still be incorporated within the MC-SDDP framework \citep{Lhndorf2013OptimizingTD}, which supports a broad class of autoregressive and moving average processes \citep{Infanger1996, 8024054, GUEVARA2024110084}.

Although convexity is lost due to the problem's mixed-integer nature and optimality is no longer guaranteed, finite convergence is proven \citep{papavasiliou2018application}. It will be shown that the quality of SDDP solutions compares well with those of direct MILP formulations. This is the only practical algorithm capable of solving the PSTP problem at a realistic planning scale \citep{Füllner_Rebennack_2022}.

\section{Data Framework and Scenario Construction}
\label{section4}

\begin{figure}[ht]    \centering\includegraphics[width=0.70\textwidth]{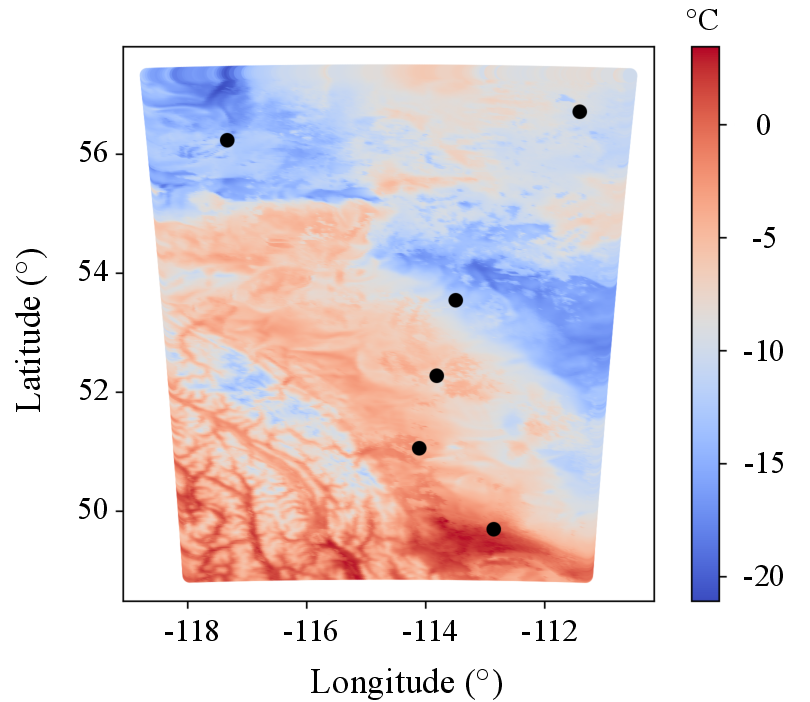}
    \caption{Heat-map spanning AESO-6 produced from a single partition (1 hour).}
   \label{fig:Area}
\end{figure}

\noindent Weather data is downloaded from the Canadian Surface Prediction Archives (CaSPAr) \citep{Mai_Kornelsen} for a whole year (Jan 2022 - Jan 2023). Specifically, High-Resolution Deterministic Prediction System (HRDPS) data is used. Those provide hourly temporal (up to 48h predictions) and 2.5$km^2$ spatial resolution in most of Canada and North of the USA for various parameters, such as air temperature, wind speed and direction, and downward short-wave (direct) and long-wave radiation (diffuse) flux. The data is filtered and spatially cropped to meet the desired area encompassing the test case jurisdiction, as shown in Fig. \ref{fig:Area}. 

Elevation data is downloaded through the Google Maps application programming interface \citep{GoogleMaps} using the exact coordinates of the spatiotemporal data. The availability of such correlated data allows us to map our problem geospatially for more reliable planning. For optimal processing and parallel calculation of the extensive mesh data, the Dask-Dataframe package in Python is used to load the data in 8760 partitions representing hourly snapshots of data of the year; each partition holds a Dataframe representing a mesh of point predictions corresponding to the longitudes and latitudes of locations. 

\subsection{Dynamic Thermal Rating}

Dynamic Thermal Rating (DTR) calculation followed the IEEE-738 standard \citep{IEEEDTR} for every point in the mesh for every hour. The maximum allowed conductor temperature is assumed to be 100°C, and the assumed line properties are those of a typical aluminum conductor steel reinforcement following Alberta Electric System Operator (AESO) standards \citep{ISO502}. The resulting data from the DTR calculation is saved for a subsequent time-aggregation step, and the DTR for the specific properties of existing and candidate lines is calculated after the representative days are determined. 

\subsection{VRES Power Output}

The calculation of solar power output using direct and diffuse irradiation depends on the type of solar panel and tracking system, orientation, and other factors \citep{PV_Education}. For simplicity, a typical solar module with $\eta$ = 22\% efficiency is assumed \citep{Ghzizal_2022}. Also, the effect of air temperature and wind speed on the output of the PV module is ignored, and the most common tracking system, the horizontal single-axis tracker, is assumed \citep{Ghzizal_2022}. With these assumptions and using the downloaded irradiation data, the hourly solar power output can be calculated (\ref{65}) and (\ref{66}) \citep{Sandia043535}:

\vspace{-10pt}
\begin{align}
    \text{POA}_t & = \frac{D_t^{\text{dirc}} + D_t^{\text{diff}}}{cos(|90 - \theta_t|)} \tag{65} \label{65} \\
    Z_t &  = \text{POA}_t \times \eta \tag{66} \label{66}
\end{align}

$D^{\text{dirc}}$ is the direct downward solar flux, $D^{\text{diff}}$ is the diffuse downward solar flux, POA is irradiance on the plane of incidence, $\theta_t$ is the angle of the sun at hour $t$. $\eta$ is the efficiency of the solar module, and $Z$ is the final solar power output at time $t$. The calculation of wind power output is even more straightforward: choosing a typical wind turbine \citep{Goldwind} and using its power curve to determine its production from the downloaded wind speed data \citep{burton_wind_2011}.

\begin{figure}[!htbp]
    \centering
\includegraphics[width=0.7\textwidth]{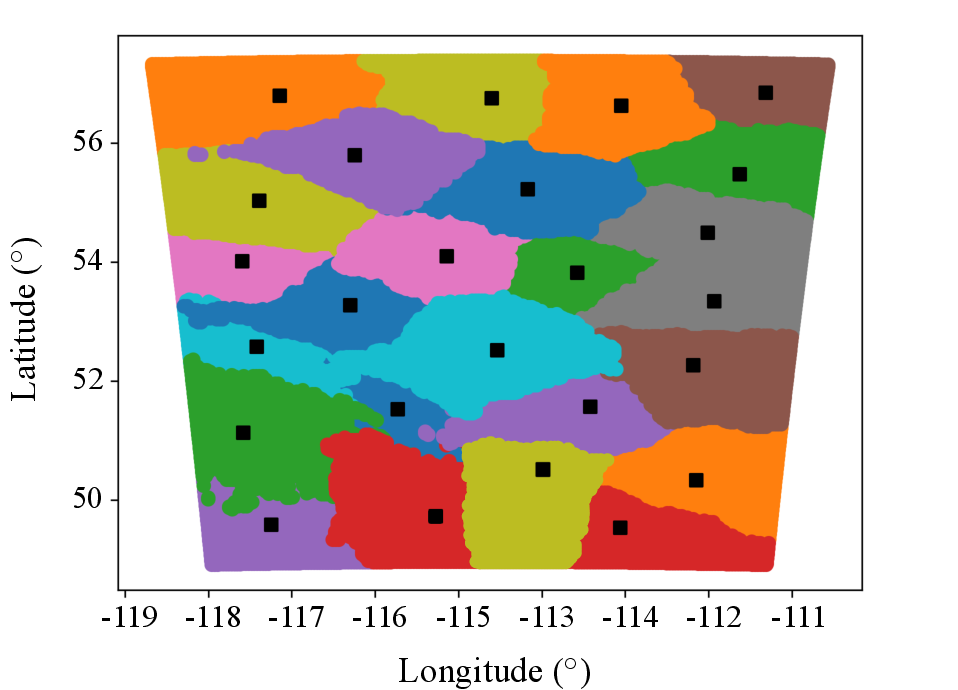}
    \caption{Clustered locations with their medioids}
   \label{fig:REclust}
\end{figure}

The vast number of data points heavily increases the computational burden. Thus, the AESO-6 test case uses a reduced set of zones as seen in Fig. \ref{fig:REclust}. Zone number reduction is done by clustering the yearly time series data of each zone using Dynamic Time Wrapping (DTW) \citep{Berndt1994UsingDT} into $k$ clusters, then extracting the medoid zone to represent the output of all zones within the same cluster. The area available for each medoid zone is  $2.5km^2$ (data resolution) multiplied by the number of cluster data points the medoid represents. Then, every medoid zone is associated with the nearest bus, which determines the injection point of the VRES installed in the zone. Fig. \ref{fig:REclust} shows the span covered by 25 representative zones.

In practice, many areas covered by the data may not be suitable for wind and solar allocation and are subject to geographical and regulatory limitations. The precise boundaries are not defined as such in this test case, but the measure of reducing all the areas by a factor of 10 is taken based on Alberta’s protected area map \citep{CPAWS2024}. For the large test case, a numerical experiment is performed in Section \ref{section6.3} to evaluate the impact of the number of zones on the solution quality and computational burden to determine a suitable data resolution.

\subsection{AESO-6 test case}

\begin{figure}[ht]
    \centering
\includegraphics[width=0.90\textwidth]{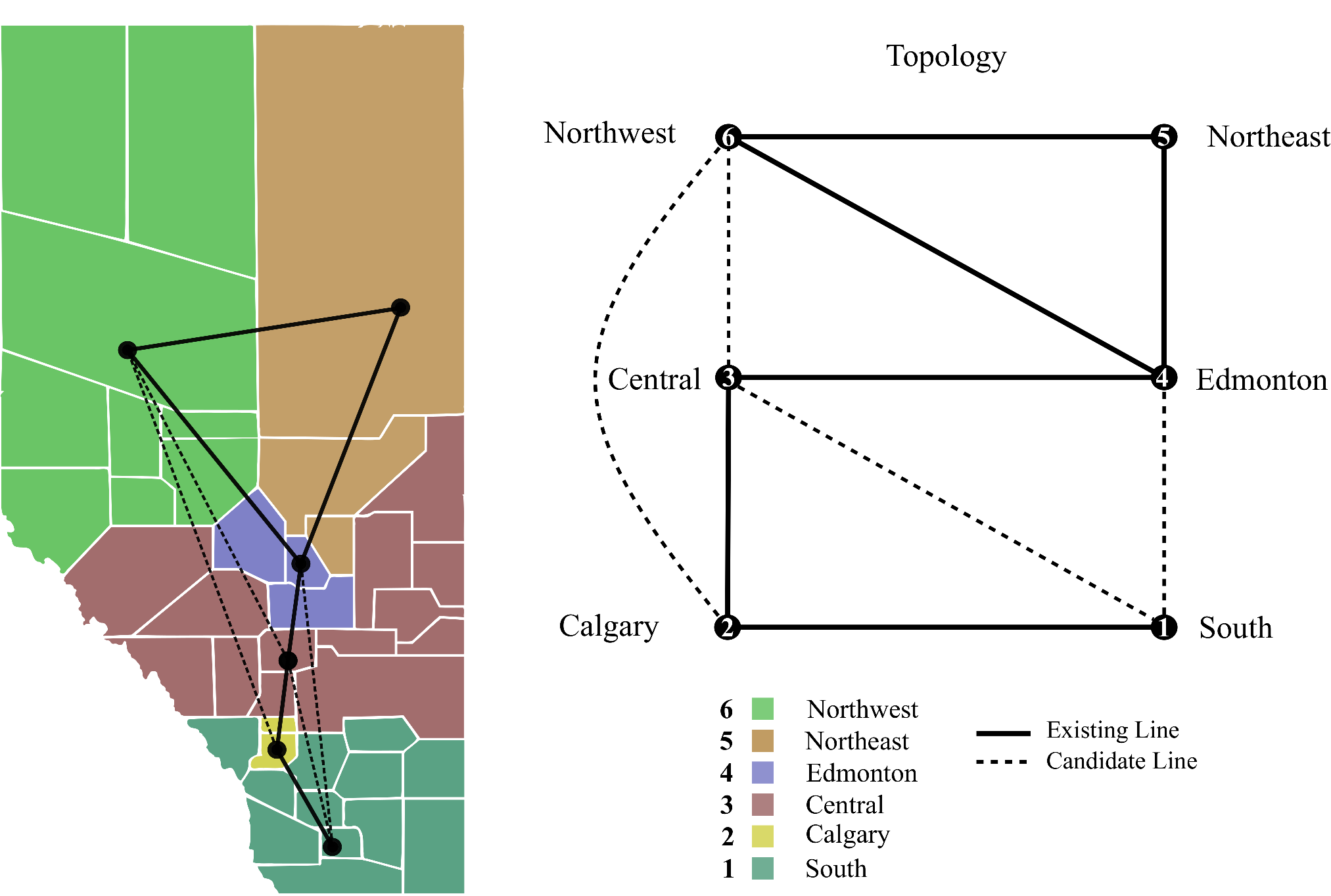}
    \caption{AESO-6 system projected on AESO’s planning areas.}
   \label{fig:AESO6}
\end{figure}

The 6-bus test case used in this paper is adapted from Alberta’s power system, selecting one bus to roughly represent each load region, aggregating the generators of the buses in the regions, and transmission lines as shown in Fig. \ref{fig:AESO6}. The candidate transmission paths in Fig. \ref{fig:AESO6} are chosen to mitigate existing congestion as indicated by the Alberta Transmission Utilization map \citep{AESO2024}. To determine the static rating of the existing lines, it is first assumed that all the lines have infinite capacity and solve the optimal power flow problem for each region’s average load of the past year. The lines' resulting flow determines the existing lines' capacity for each Right of Way (RoW). The existing products or properties of the line that achieve that capacity based on AESO static rating standards are then determined \citep{ISO502}. Candidate lines in existing RoWs exhibit the same property as the existing lines, whereas candidate lines in new RoWs exhibit the property of the line with the largest capacity in the network. The installation costs of new lines are all driven from historical project costs by AESO \citep{AESO2024cost}. Tables \ref{tab:ExCap} and \ref{tab:TransRow} in \ref{app:tables} show the final relevant properties of the test case used, and the test case data is available at \citep{AESO1446}.

\subsection{Scenario Generation}

\subsubsection{Intrastage Scenarios}

\begin{figure}[ht]
    \centering\includegraphics[width=0.70\textwidth]{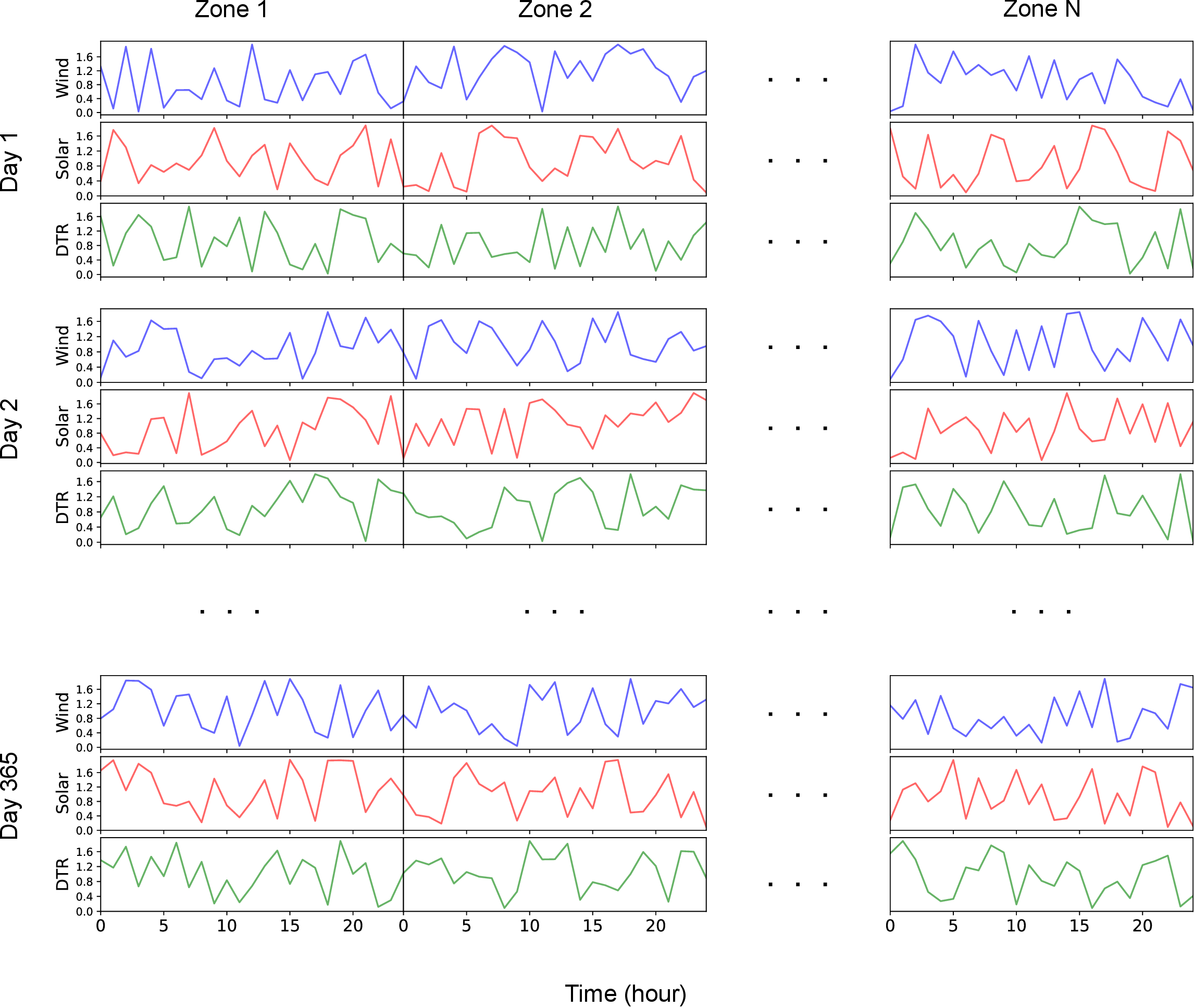}
    \caption{An illustration of array preparation for multivariate clustering}
   \label{fig:DTW}
\end{figure}

The short-term or intrastage scenarios include wind production, solar production, DTR, and load profiles. The load profiles for the test case are downloaded from AESO’s website \citep{AESO22}, which provides aggregate loads coinciding with the regional division of the test case and the location of buses. To create the reduced set of short-term scenarios, the 24 hours are split, and the daily profiles of all zones $z$ are linked for each random variable $\zeta$ sequentially, creating 365 two-dimensional arrays of size $N_\zeta\times24z$ as shown in Fig. \ref{fig:DTW}, where $N_\zeta$ is the number of random variables. Then, multivariate clustering of the days is performed using Multi-dimensional Dynamic Time Wrapping \citep{ShokoohiYekta2015OnTN}, and the medoids of the clusters are found. Those medoids would represent the days used for the short-term scenarios to capture representative days of the whole set without losing any spatiotemporal correlations.

\subsubsection{Out-of-Sample Analysis}

\begin{figure}[ht]
    \centering
    \includegraphics[width=0.70\textwidth]{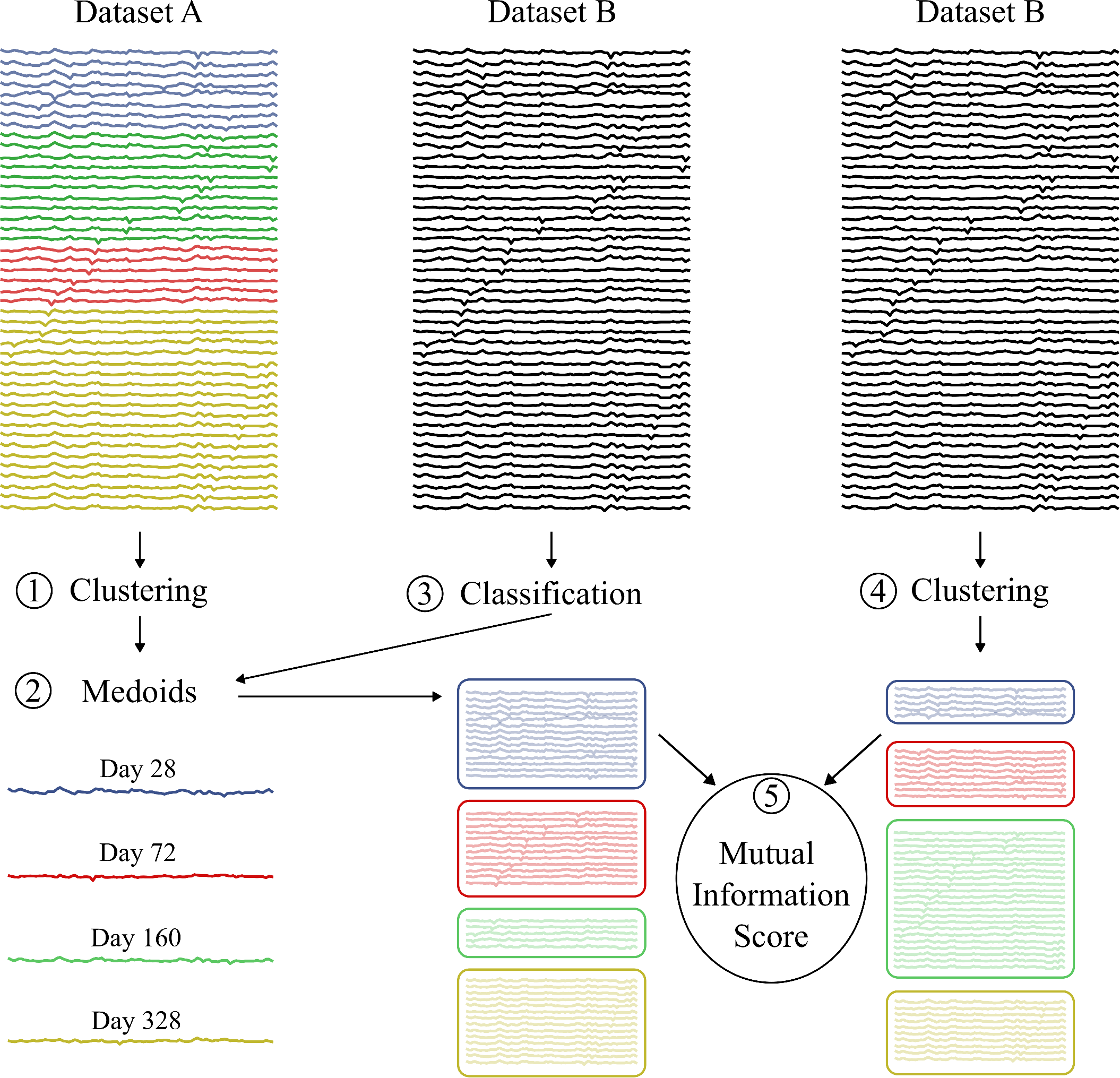}
    \caption{An illustration of the out-of-sample analysis test of short-term scenarios}
   \label{fig:MutInfo}
\end{figure}

To validate the clustered profiles and determine whether they sufficiently represent out-of-sample data, an approach similar to the one used in \citep{nima2023} is followed. First, the same clustering method described earlier is applied to a validation set of historical data, including all the years from 2020 to 2023. Then, using the pairwise DTW \citep{Berndt1994UsingDT} to measure the distances, the validation set data points are grouped using the medoids of the original dataset as fixed centroids. After that, the similarity between the validation set clusters and the clusters created using the original set medoids is compared by using Mutual Information (MI) \citep{Romano2016}. Fig. \ref{fig:MutInfo} illustrates how the MI score was obtained by comparing the two clustering methods. Dataset A in Fig. \ref{fig:MutInfo} represents the original data from which the medoids are extracted, representing the short-term scenarios for the model. Dataset B represents the validation data; its data points are clustered twice, once by classifying them using the medoids of A and again using unsupervised clustering using DTW, as mentioned previously.

Table \ref{tab:3} shows the MI scores obtained where the normalized MI (NMI) and Adjusted MI (AMI) values are calculated using the averaging method in \citep{Vinh2009}. The NMI score in Table \ref{tab:3} confirms a strong agreement between the two clusters, as 1 indicates a perfect alignment and 0 indicates no agreement. The adjusted MI score shows that the clusters are well aligned even after accounting for the chance of random overlap. Generally, adjusted MI scores between 0.7 and 0.8 are considered good alignment indicators \citep{Romano2016}. As the extracted profiles classify dataset B into clusters with high similarity to the clusters obtained from the unsupervised clustering of the same dataset, it suggests that while the generalization is imperfect, the four profiles adequately capture the overall behavior of the uncertain parameters \citep{nima2023}. 

\begin{table}[ht]  
\caption{Mutal Information Scores}
  \label{tab:3}
  \centering
  \small
  \begin{tabular}{lccc}
\toprule
Metric    & MI & Normalized MI & Adjusted MI \\
\midrule
Score & 1.007 & 0.7520& 0.7497 \\
\bottomrule
  \end{tabular}
\end{table}

\subsubsection{Interstage Scenarios}

While the time aggregation method described previously constitutes the noise in every node, its magnitude changes by a scaling factor depending on the system’s state at each node. The system's state in this model affects the short-term scenarios and other long-term random variables that only change between stages and are not part of the noise. In this model, the long-term random variables are fuel and technology allocation prices, which differ at every node but do not fluctuate hourly. To create the scenarios for this test case, information is extracted from several sources and consolidated to represent three system states per stage. Those are labeled Baseline (B), Moderate (M), and Optimistic (O). Tables \ref{tab:4} and \ref{tab:5} summarize the system states of the MC-SDDP model and the equivalent scenario path of the Monolithic model, respectively, for interstage scenarios.

Demand growth scenarios are extracted from AESOs 2024 long-term energy outlook \citep{Orano2024}, where three peak internal load scenarios are provided, including 1- historical load growth (B), 2- developing electrification (M), and 3- high electrification (O). Technology and fuel cost scenarios are extracted from the most recent NREL ATP report \citep{NRELATB}; those are 1- Conservative (B), 2- Moderate (M), and 3- Advanced (O). The climate scenarios chosen coincide with the five Intergovernmental Panel on Climate Change (IPCC) SPP \citep{ipcc2021ar6}, which are 1-SSP5 (B), 2- SSP3 (M), and 3-SSP1 (O). The data is extracted from \citep{CanadaCMIP62024}, which provides Coupled Model Intercomparison Project Phase 6 (CMIP6) predictions of temperature and wind speed changes for the province of Alberta. In each stage, long-term scaling factors have exhibited different values. For example, in stage two, the baseline state value for the load is 1.1, and in stage three, the value is 1.2. This value is then multiplied by the short-term load fluctuations. 

\begin{table}[t!]  
\caption{MC-SDDP Stage States}
  \label{tab:4}
  \centering
  \small
  \begin{tabular}{ccccc}
\toprule
& State      & Tech \& Fuel & Climate & Load  \\
\midrule
1 & Baseline   & Conservative & SSP5   & Historical\\
2 & Moderate   & Moderate     & SSP3   & Developing \\
3 & Optimistic & Advanced     & SSP1   & High \\
\bottomrule
  \end{tabular}
\end{table}

\begin{table}[t!]  \caption{Monolithic Scenario Tree Paths}
  \label{tab:5}
  \centering
  \small
  \begin{tabular}{cccccccccc}
\toprule
Scenario Path & 1 & 2 & 3 & 4 & 5 & 6 & 7 & 8 & 9 \\
\midrule
Stage 2  & O & O & O & M & M & M & B & B & B \\
Stage 3  & O & M & B & O & M & B & O & M & B \\
\bottomrule
  \end{tabular}
\end{table}

\subsection{Transition and Operation Costs}

The required transition and operation cost data are extracted from the most recent NREL ATP report \citep{NRELATB}. The DTR device is assumed to be the commercially available \citep{LAKI_POWER}. While only one critical zone per branch is used to calculate DTR, multiple sensors might be placed on a single line in practice. Some studies attempt to determine the critical span between sensors \citep{Rui_Sahraei}. It is a challenging process and is outside the scope of this study. For this test case, it is sufficient to assume that a sensor has to be placed every 3km of the line, which is generally a good arrangement based on sag monitoring devices \citep{ Karimi2018}, making the capital cost of DTR a function of the line length. Load-shedding and curtailment costs are subjective and differ from case to case. VOLL is assumed to be \$100/MW following \citep{AESO_2019}. Because VRES curtailment cost estimation varies and is governed by different policies, a tentative penalty is set at a value slightly lower than the lowest fuel cost based on the idea that curtailment should occur when the cost of avoiding it equals its marginal value \citep{KLINGEJACOBSEN2012663}.  After defining the unit costs, calculations are performed to parameterize the transition costs at each stage. Equations (\ref{67}) and (\ref{68}) show how the cost for the new gas generation $I^G$ is calculated:

\vspace{-5pt}

\begin{align}
    F_y & = F^a \times Y^s \times R_y  \tag{67} \label{67} \\ 
    I^G_{y,s}  & = (\text{CapEx} + F_y) \zeta^{I}_{y,s} \tag{68} \label{68}
\end{align}

$F_y$ is the fixed operation and maintenance cost (FOM) at stage $y$; it is the multiplication of the annual FOM ($F^a$), the number of years implied per stage ($Y^s$) (e.g., if the number of planning stages is two and the planning horizon is ten years, $Y^s$ becomes five years), and the remaining stages of the planning horizon from stage $y$ ($R_y$). The sum of costs (CapEx + $F_y$), where CapEx is the p.u. fixed cost of the technology is multiplied by $\zeta^I_{y,s}$, which is the realization of price change for every long-term scenario determined based on the aforementioned NREL scenarios \citep{NRELATB}.

\section{Monolithic Solution and Analysis}
\label{section5}

\noindent First, the problem is modeled monolithically on Pyomo \citep{hart2011pyomo} using the AESO-6 system. In this test case, transition decisions are made over three stages; each stage represents five years with three Markovian states per stage, as shown in Fig. \ref{fig:MCchain} and Table \ref{tab:4}. Four representative days are chosen using the method described in subsection D to represent the noise profiles in each state for solar and wind output, DTR, and load. 25 Wind and solar candidate zones are chosen using the method described in subsection B. Each variable varies per state, amplified or reduced by the state's scaling factor. Short-term scenarios were assigned probabilities proportional to their cluster size (weighted medoid representation). The long-term outlook scenarios usually do not have associated probability and are treated as ``equally valid," removing any potential bias \citep{Millett2009ShouldPB}. Probabilistic socioeconomic projections require a rigorous assessment of a large set of socioeconomic variables and are prone to bias in the estimation \citep{Meinshausen2020TheSS}. Producing or choosing a specific extension to the IPCC SPP is outside the scope of this work. Thus, the long-term scenarios were assigned the unbiased equal probabilities assumed by their reports \citep{ipcc2021ar6, NRELATB, Orano2024}.

In a scenario tree, this equates to nine long-term scenarios and four short-term scenarios per node. This equate to 144 possible scenario realizations (2 stages $\times$ four short-term scenarios $\times$ nine long-term scenarios), or in MC-SDDP it equates to 144 possible forward samples (referred to as scenarios in MC-SDDP context \citep{dowson2021sddp}), since there is one node in the first stage, three nodes in the second and third stages, and only one of the four short-term scenarios is realized in each node forward pass ($1 \times (3 \times 4) \times (3 \times 4) = 144$).

\subsection{Results}

\begin{figure}[!htbp]
    \centering
\includegraphics[width=0.70\textwidth]{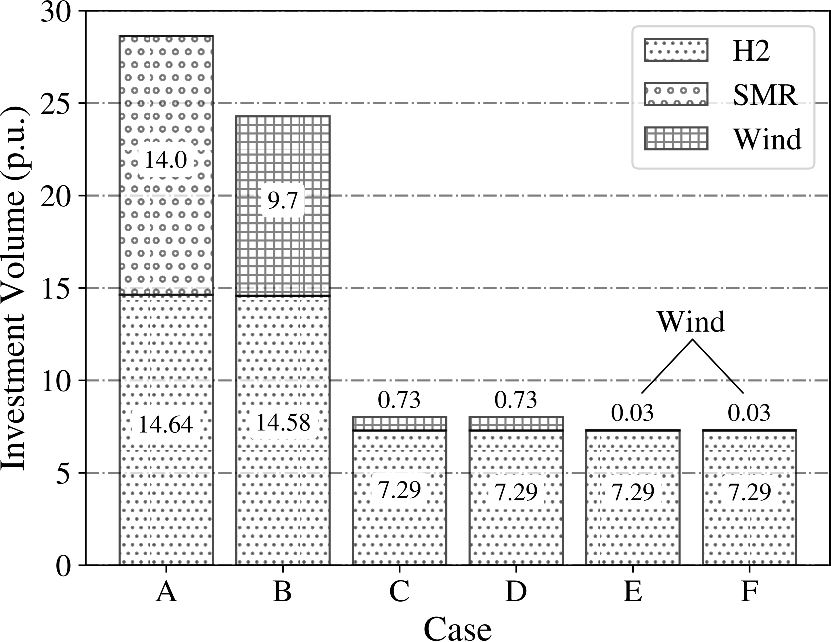}
    \caption{Nominal capacity of allocated resources in the first stage.}
   \label{fig:Inv1}
\end{figure}

\begin{figure}[!htbp]
    \centering
    \includegraphics[width=0.70\textwidth]{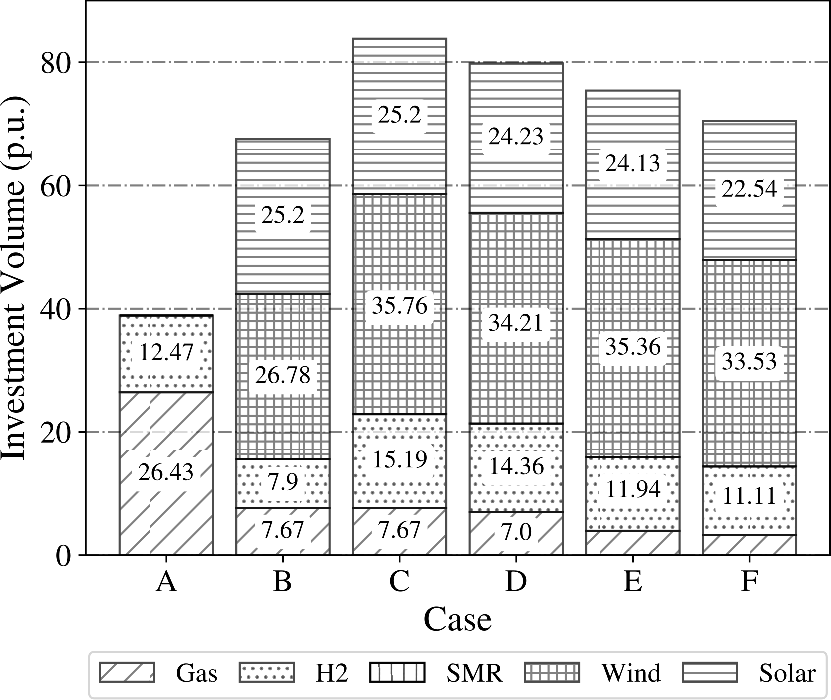}
    \caption{Nominal capacity of allocated resources in the second stage.}
   \label{fig:Inv2}
\end{figure}

\begin{figure}[!htbp]
    \centering
\includegraphics[width=0.70\textwidth]{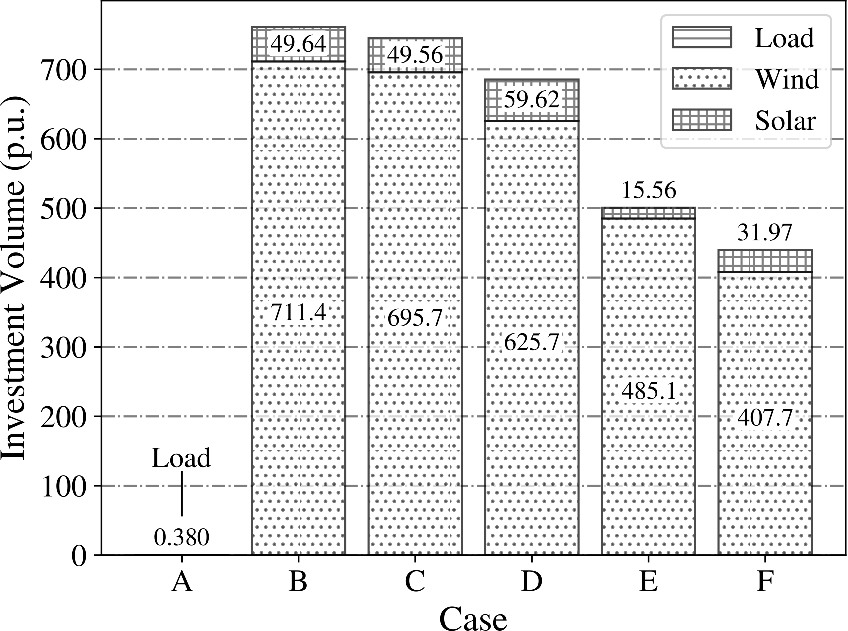}
    \caption{Volume of total curtailment of RE and load for the whole planning horizon.}
   \label{fig:curtailment}
\end{figure}

The produced MILP problem is solved using the Gurobi solver \citep{gurobi}  for several test cases with different combinations of planning factors at a time. Table \ref{tab:6} shows the factors included in each test case with the optimal cost in each case. Table \ref{tab:7} shows each case’s binary planning factor locations. Only the decisions for the first two stages are shown because a three-stage MSP implies two here-and-now decisions in the first and second stages and two wait-and-see decisions in the second and third stages. Fig. \ref{fig:Inv1}, Fig. \ref{fig:Inv2}, and Fig. \ref{fig:curtailment} show the allocated capacities and predicted curtailments in the first and second stages of the planning horizon. Note that the second stage transition decisions refer to the ones explicitly made in scenario three just for demonstration, as there are three possible states in the second stage.

\begin{table*}[ht]  \caption{Monolithic AESO-6 bus Results}
  \label{tab:6}
  \centering
  \small
  \resizebox{\textwidth}{!}{%
  \begin{tabular}{cccccccccccccc}
      \toprule
      \textbf{Case} & \multicolumn{11}{c}{\textbf{Planning Factors}} & \multicolumn{2}{c}{\textbf{Optimal Cost}} \\
      \toprule
      Case & Gas & H2 & SMR & Solar & Wind & CCS & Batt.      
 & Pump  & Line & DTR & SSSC & 1st Stage (\$b) & Overall (\$b)  \\ 
      \midrule
      A & \checkmark & \checkmark & \checkmark & - & - & - & - & - & - & - & - & 14.70 & 40.74 \\
      B & \checkmark & \checkmark & \checkmark & \checkmark & \checkmark & - & - & - & - & - & - & 3.690 & 15.87  \\
      C & \checkmark & \checkmark & \checkmark & \checkmark & \checkmark & \checkmark & - & - & - & - & - & 1.095 & 15.75 \\ 
      D & \checkmark & \checkmark & \checkmark & \checkmark & \checkmark & \checkmark & \checkmark & \checkmark & - & - & - & 1.716 & 15.49 \\
      E & \checkmark & \checkmark & \checkmark & \checkmark & \checkmark & \checkmark & - & - & \checkmark & \checkmark & \checkmark & 0.997 & 13.94 \\
      F & \checkmark & \checkmark & \checkmark & \checkmark & \checkmark  & \checkmark & \checkmark  & \checkmark  & \checkmark  & \checkmark  & \checkmark & 1.613 & 13.56 \\
      \bottomrule
  \end{tabular}
}
\end{table*}

\begin{table}[ht] 
\caption{Asset location (RoWs/Buses)}
  \label{tab:7}
  \centering
  \scriptsize
  \begin{tabular}{cccccccc}
    \toprule[1pt]
    Stage & Case &  C & D & E & F   \\ 
    \midrule[1pt]
   1&  CCS   & \{19\} & \{19\}    & \{19\}  & \{19\} \\
   2&  CCS   & - & -   & - & - \\
    \midrule
   1& Lines   & -         & -            &  -        & - \\
   2& Lines   & -         & -            &  \{10\}       & \{10\}\\
    \midrule
   1& DTR     & -         & -            &   \{1,2,3,6\}        &   \{1,2,3,6\}          \\
   2& DTR     & -         & -            &  -        &      -      \\
    \midrule
   1& SSSC    & -         & -            & \{1, 2, 4, 5, 6\}     & \{1, 2, 4, 5, 6\}  \\
   2& SSSC    & -         & -            & \{4\}     & \{4\}  \\
    \midrule
   1& Battery & -         & - &  -        &      -       \\
   2& Battery & -         & \{4, 6\}  &  -       &      \{4, 6\}      \\
    \bottomrule[1pt]
  \end{tabular}
\end{table}

From the results, several critical insights can be obtained. The first is that the more factors are included in the model, the lower the optimal overall cost is, as seen in Table \ref{tab:6}. In Case A, where only rotary generation is included, the deferral of gas generation and some hydrogen turbine (H2) unit allocation to the second stage can be noticed in Fig. \ref{fig:Inv2}. Also, some load shedding is predicted, as seen in Fig. \ref{fig:curtailment}, due to the limited planning factors, heavily increasing the overall cost in Case A. When VRES options are included in Case B, there is a notable reduction in rotary generation allocation, as renewables enter the pool of planning factors, and load-shedding is eliminated, drastically decreasing the overall cost. Solar generation is allocated to the second stage, where its per unit cost drops by 32\% in the displayed scenario (compared to 8\% for wind), making it economical to defer its investment to the second stage. It is also noted that wind curtailment is exceptionally high, contributing to the overall cost. In Case C, when the retrofitting option becomes available, the allocation of new H2 generation and wind units is deferred to the second stage because of the retrofitting decision made on the gas unit 19 as seen in Table \ref{tab:7}, which makes it viable in meeting annual emission targets. In Case D, a reduction in wind curtailment can be noticed when storage options are included. In Case E, storage options are omitted, and, instead, transmission reinforcement options are allowed, namely, installation of DTR, new line, and FACTS. The total cost and VRES curtailment in Case E are lower than those in Case D, suggesting that modular network management elements are more effective in reducing costs than storage systems in this particular case. As seen in Table \ref{tab:7}, in case E, DTR is only allocated on several lines at the first stage, while SSSC allocation is spread out between the stages. Namely, lines 1, 2, 4, 5, and 6 have 6, 8, 3, 4, and 6 SSSC units allocated in the first and second stages, respectively. In the second stage, line 4 has three SSSC units allocated. For transmission lines, line 10 (Edmonton to South) is installed in the second stage only in scenario 3. Finally, in Case F, where all the factors of this model are included, the first-stage optimal cost is higher than that in the previous case. However, this reduces the optimal overall cost due to the decreased curtailment and operational costs. It demonstrates the advantage of multistage planning, where a substantial initial allocation cost reduces long-term costs by proactively managing future risks. 

\subsection{Stochastic Solution Quality}

\begin{figure}[!ht]
    \centering
\includegraphics[width=0.7\textwidth]{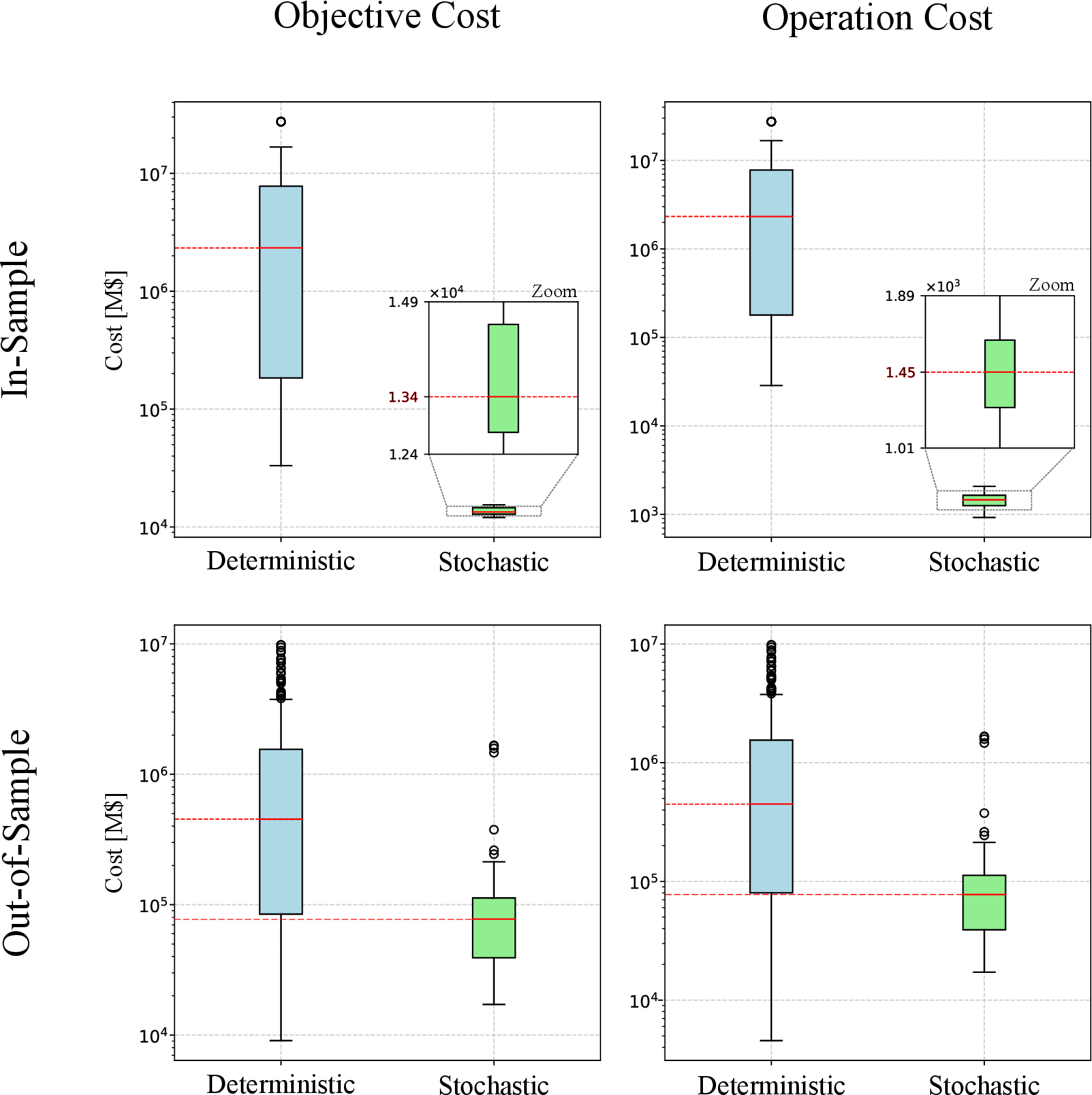}
    \caption{In-sample and out-of-sample scenario-wise ex-post comparison of deterministic and stochastic solution performances}
   \label{fig:Voss}
\end{figure}

To assess the quality of the stochastic solution, we compute the Value of the Stochastic Solution (VoSS) by first solving a deterministic version of the problem using the expected values of uncertain parameters. The resulting decision vector is then evaluated within the complete stochastic framework to obtain the expected result of using EV (EEV). Comparing this with the original stochastic model's Recourse Problem (RP) solution yields VoSS, as shown in Eq. \ref{69}. Beyond this, we conduct a scenario-wise ex-post evaluation by applying both EEV and RP solution vectors to 144 representative in-sample realizations, followed by 10,000 out-of-sample realizations generated from historical load data. As seen in Fig. \ref{fig:Voss}, in all cases, investment decisions are fixed while operational costs are re-optimized, enabling the comparison of total and operating costs. Investment costs vary across realizations due to the change in cost across long-term scenarios.

\begin{align}
VSS = EEV - RP = 15.73 - 13.56 = \$ 2.17 B \tag{69} \label{69}
\end{align}

Notice how in the in-sample and out-of-sample scenarios, the deterministic solution underperforms significantly, primarily due to operational cost and elevated load-shedding and curtailment penalties, underscoring the practical benefits of the stochastic approach. The difference remains stark in the out-of-sample.

\section{SDDP solution}

\label{section6}

\subsection{AESO-6 Comparision}

\noindent To verify the SDDP’s accuracy, the problem is solved using SDDP on the same AESO-6 test case, including all the factors (Case F). The Julia package SDDP.jl implementation of the algorithm is used \citep{dowson_sddp.jl}. This test case is solved both monolithically and using the SDDP algorithm. The convergence criteria or stopping rule is set to bound stalling in all trials, further explained in  \citep{dowson2021sddp} \citep{shapiro2009sddp}. The algorithm ends if the lower bound fails to improve by more than a pre-determined tolerance (1e-4) after 25 consecutive iterations. The upper bound for the algorithm is not exact. It is obtained by performing a Monte Carlo simulation on the obtained policy (deterministic first-stage problem). Thus, it is only used as a sanity check by inspecting whether the lower bound falls within the upper bound's confidence
interval \citep{homem2011sampling}.  Both models are solved on the University of Calgary HPC platform using Intel Xeon Gold 6148 2.40GHz CPUs and 180GB of memory. Table \ref{tab:8} shows the results of solving the monolithic and SDDP models. 

\begin{table}[ht]
  \caption{Experiment results}
  \label{tab:8}
  \centering
  \scriptsize
    \begin{tabular}{lcccc}
    \toprule
    \textbf{Model} & \multicolumn{2}{c}{Monolithic}   & \multicolumn{2}{c}{SDDP}    \\
    \midrule
    \textbf{Physical Cores} & \multicolumn{1}{c}{1}   &  \multicolumn{1}{c}{40} &  \multicolumn{1}{c}{1} &  \multicolumn{1}{c}{40}\\ 
    \textbf{Parallel Scheme} & \multicolumn{1}{c}{-}   &  \multicolumn{1}{c}{Parallel MIP} &  \multicolumn{1}{c}{-} &  \multicolumn{1}{c}{Serial SDDP}\\ 
    \textbf{Wall Time [m]} & \multicolumn{1}{c}{977.2}   &  \multicolumn{1}{c}{59.612} &  \multicolumn{1}{c}{126.51} &  \multicolumn{1}{c}{19.730}\\ 
\midrule 
\textbf{Variables} & \multicolumn{2}{c}{338185}   & \multicolumn{2}{c}{7823}    \\
    \textbf{Constraints} & \multicolumn{2}{c}{729223}   & \multicolumn{2}{c}{20951}    \\
    \textbf{Scenarios} & \multicolumn{2}{c}{9}   & \multicolumn{2}{c}{144}    \\
    \textbf{Subproblems} & \multicolumn{2}{c}{-}   & \multicolumn{2}{c}{7}    \\
    \bottomrule
  \end{tabular}
\end{table}

Note that the number of scenarios in the SDDP column in Table \ref{tab:8} refers to the sample paths in the Markov Chain, and the number of variables and constraints is for each subproblem. Moreover, ``Serial SDDP" refers to a synchronous parallel process. The parallel MIP algorithm executes parallel instances exploring independent frontier nodes of the branch and bound (B\&B) tree concurrently. In contrast, Serial SDDP solves copies of the MC-SDDP problem in separate processes, sampling different scenarios in each process and sharing the discovered cuts with the master process.

As seen in Table \ref{tab:8}, the SDDP algorithm outperforms the monolithic solution in single-core and 40-core cases, converging to the same optimal solution of \$13.56b. The single-core performance of SDDP warrants particular attention as it is comparable to the 40-core performance of the monolithic solution. It is important to note that this small test case has low granularity for SDDP, meaning that it is heavily bottlenecked by data communication. It is also important to note that the Parallel MIP used in the monolithic solution does not scale linearly and has unpredictable behaviour as (B\&B) trees can be deep and thin (unbalanced) and hard to parallelize  \citep{gurobi}.

The main bottleneck of the ``Serial SDDP" is that the parallel sub-processes are halted by the master process, waiting for all other processes to share their cuts before moving on to the next iteration. This explains the sub-linear time scaling over the number of cores. Asynchronous SDDP can speed up iterations, immediately sharing discovered cuts between workers. It has the caveat of a long set-up time at the start of the algorithm to establish the communication protocols. Thus, it is more suitable for larger problems where the algorithm spends a significantly longer solving the subproblems. The large-scale test cases in Section \ref{section6.4} are all intractable monolithically, yet, as will be shown, converged by using SDDP.

\subsection{AESO-144 test case}
\begin{figure}[h!]
    \centering
\includegraphics[width=0.720\textwidth]{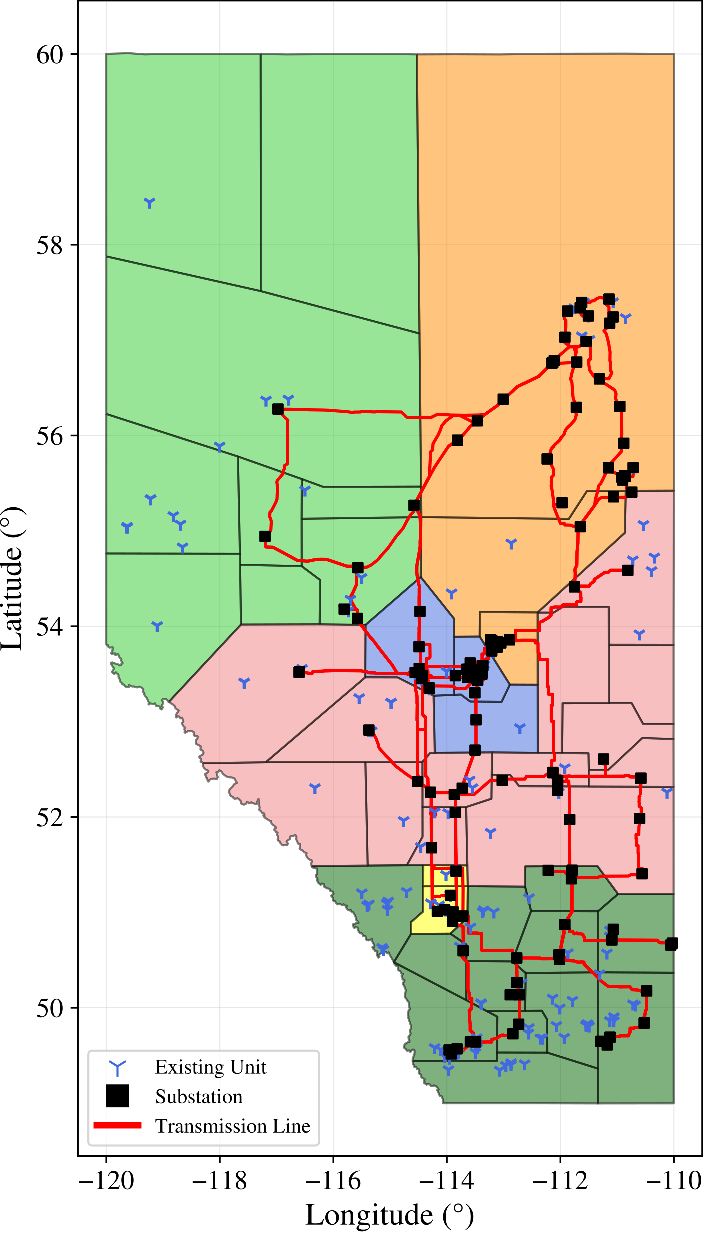}
    \caption{AESO-144 test case}
   \label{fig:AESO144}
\end{figure}

At system-level planning, the AESO often only includes the 240kV and 138kV corridors; this is followed by creating a realistic planning test case to showcase the algorithm’s scalability. Reducing the AESO system to the corridors above results in a 144-bus system with 217 branches. Fig. \ref{fig:AESO144} shows the topology of this network. It is important to note that the bus count in this framework is a poor proxy for complexity, given the total number of binary variables, uncertainty layers, and the resolution of renewable geospatial endowment. Moreover, unlike previous work,  this test case sticks to the realistic planning procedure of the ISO without any reduction and remains larger than any network in comparable work. This test case’s geographical and system information is created by extracting the data from \citep{AESO2024, NRGstream} and is available in \citep{AESO1446}. The same steps used for the AESO-6 test system are applied to parameterize the AESO-144 test case.

As seen in Fig. \ref{fig:AESO144}, the branches do not extend between buses in straight lines in a real test case. This is an important issue that is accounted for in DTR calculation, as the azimuth of the line plays a significant role in convection heat loss. The candidate lines are determined based on the utilization rate. Any lines with a 75\% utilization rate are of concern to AESO and suggest that the system is under-built \citep{AESO2023}. Those lines are determined from \citep{AESO2024} and chosen as candidates for new-line installation, resulting in 32 candidate lines, whereas SSSC and DTR devices are allowed for all branches. A single candidate hydro-pump system is set according to a proposed future project in Alberta \citep{CanyonCreekPumpedStorage}. Candidate battery systems and new rotary generation options are allowed on every bus. The planning horizon is set to 20 years (2025 - 2045). In the small test case (AESO-6), three out of five SSP narratives \citep{ipcc2021ar6} were used to create the long-term scenarios. All five scenarios are included to scale the AESO-144 test case model further, resulting in 5 nodes per stage.

\subsection{VRES Data Resolution}
\label{section6.3}
\begin{figure}[!htbp]
\centering\includegraphics[width=0.70\textwidth]{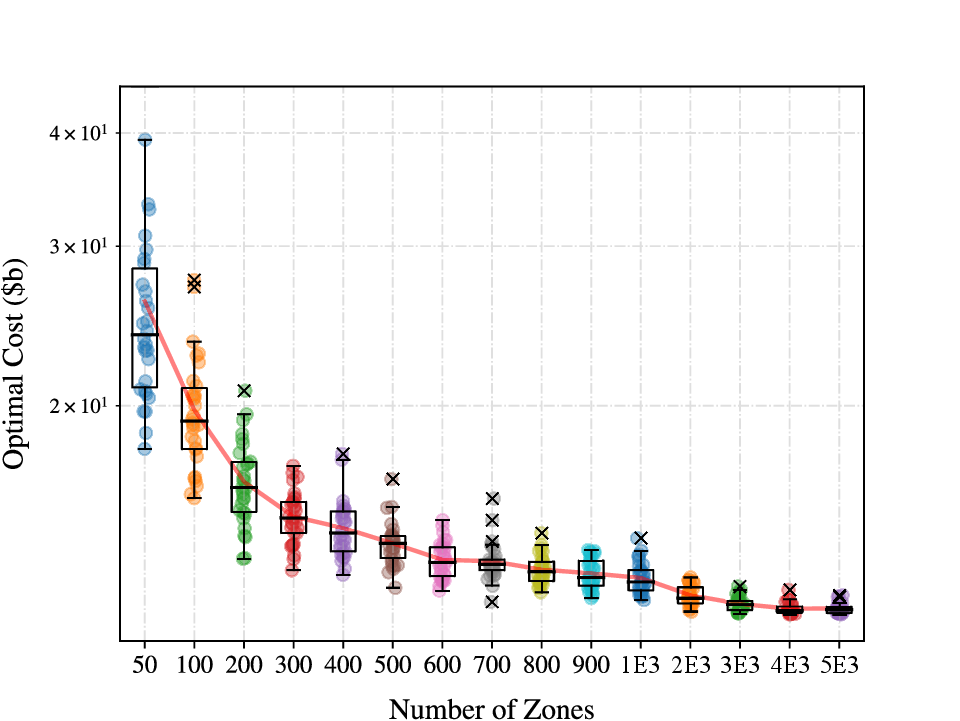}
    \caption{Solution quality based on number of zones (log scale).}
   \label{fig:ZoneTest}
\end{figure}

\begin{figure}[!htbp]
\centering
\includegraphics[width=0.70\textwidth]{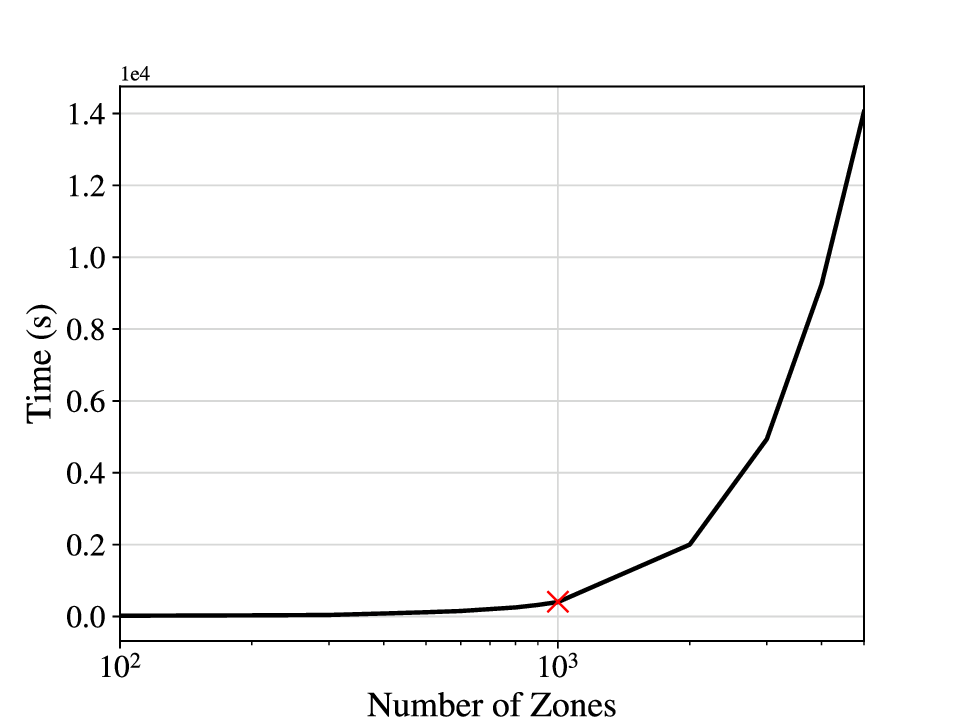}
    \caption{Solution time based on the number of zones.}
   \label{fig:ZoneTime}
\end{figure}

An experiment is designed to evaluate the trade-off between the high-resolution data and computational burden. The number of candidate zones Z on a single stage of the AESO-144 system is varied, switching off all planning factors except wind and solar energy. The zones are clustered first, as done previously, to provide a set of 5000 representative locations $L$. Suppose that $Z \subseteq L$ where $|Z|= N$. For different values of $N$, a random $Z$ is sampled from $L$ thirty times, and the model is solved each time (i.e., if \textit{N} = 100, a hundred zones are randomly sampled from $L$ thirty times, and the model is solved for each random sample). Fig. \ref{fig:ZoneTest} and Fig. \ref{fig:ZoneTime} show the optimal cost and the average computation time for the various numbers of zones.  

As seen in Fig. \ref{fig:ZoneTest},  there is an improvement in the overall cost and cost variation reduction as the number of zones increases. However, this improvement is achieved at the expense of increased solution time, as illustrated in Fig. \ref{fig:ZoneTime}. The solution time rises considerably after the 1000 zones mark (Fig. \ref{fig:ZoneTime}). It signifies the transition from a log-linear complexity to a polynomial complexity of $O(n^k)$ where $k>2$. This shift to a higher order of the computation burden growth drastically increases the theoretical worst-case runtime. Thus, to achieve the highest level of accuracy possible without an additional computational layer of complexity, 1000 zones are selected for the number of candidate solar and wind zones, representing an intermediate choice, balancing the solution quality and computational time.

\subsection{AESO-144 Results}
\label{section6.4}

The model is solved four times, varying the number of transition stages in every test while maintaining the same planning horizon (20 years). This means that only the interval at which a transition decision is made changes. In the 5-stage problem, a planning decision is made every four years. The operational cost is calculated for the four years from the moment of asset allocation. In the 20-stage problem, the transition decision is made every year. The noise profiles remain the same across the different problems. The state environment affecting the noise is specific to the particular year of each transition stage (i.e., in the 5-stage problem, the environmental states at years 2025, 2030, 2035, 2040, and 2045 apply to the noise). The results, including the total number of variables and constraints in the extensive form of the problem, are shown in Table \ref{tab:9}. The results are obtained using 80 physical cores and 1000GB of memory. The Asynchronous SDDP algorithm is used, and the termination criteria are identical to those used in the AESO-6 test case.

\begin{table}[h!]
  \caption{Experiment results}
  \label{tab:9}
  \centering
  \small
    \begin{tabular}{lcccc}
    \toprule
    No. of Stages & \textbf{2} & \textbf{5} & \textbf{10} & \textbf{20} \\
        \midrule
    \textbf{Subproblems} & 11 & 45 & 75 & 95  \\
    \textbf{Scenarios}   & 400 & 1.6$\times 10^5$ & 5.1$\times 10^{11}$ & 5.2$\times 10^{24}$ \\ 
    \textbf{Variables} & 3.1$\times 10^6$ & 6.2 $\times 10^6$ & 2.8$\times 10^7$ & 5.9 $\times 10^7$ \\ 
    \textbf{Constraints} & 5.6$\times 10^7$ & 2.2$\times 10^7$ & 5.0$\times 10^7$ & 1.1$\times 10^8$\\
    \textbf{Best Solution} [\$bn] & 33.89  & 30.50 & 30.37 & 30.30  \\ 
    \textbf{Wall Time} [m] & 560 & 2736 & 8093 & 10080 \\ 
    \bottomrule
    
  \end{tabular}
\end{table}
As seen in Table \ref{tab:9}, beyond the five stages, there is a minimal improvement in the optimal solution at the expense of an excessive increase in the solution time. The main reason the changes are not drastic when increasing the number of stages is due to the subtle changes in the system represented by the scenarios. Mainly, the sourced long-term scenarios change gradually at an almost constant rate. Unless the system experiences a considerable change in the later stages (e.g., a sudden, drastic drop in demand or wind speed from one year to another), the change in the solution would be minimal.
Nonetheless, the primary objective in addressing the extensive test case is to illustrate the scalability of the proposed solution methodology, which is shown to converge successfully. The significance of the detailed solution and the trade-off between the solution quality and computational time is more relevant to the system planner. Thus, the transition decision results are not discussed in greater detail. By aligning computational expectations with industry planning cycles, where updates are infrequent and runtime tolerance is, this work counters the assumption of intractability, which is made too quickly, typically under the constraint of solving on personal machines, expecting runtimes of only a few hours, and without exploiting HPC or decomposition techniques.

\section{Discussion}

\label{section7}

\subsection{Limitations and Challenges}

The current model expands the diversity of planning factors. However, additional aspects of PSTP present opportunities for improvement, including future planning factors such as storage systems (e.g., flow batteries, compressed-air) and flow management options like transmission line switching.

In modeling each factor, choices were made to yield significant outcomes in the final plan. Yet, specific details that could enhance the physical models, such as equipment degradation (e.g., solar panel degradation, generator decommissioning, and thermal derating), were excluded. This highlights the need for further development to ensure comprehensive model coverage.

Another detail not included in this model is the cost of transmitting power from a newly installed VRES to the associated bus. Its incorporation might involve diverse implications, potentially necessitating the addition of a fictitious node pooling of several zones together, a consideration left for subsequent implementations. 

Different tests and conditions show that the proposed model has adequate scalability for the current system size. The gap takes more than one day to reduce to the set tolerance, even with parallelism and high-performance computing. If the problem size becomes larger regarding network or variables, then the sub-problem size would be much larger, and the solution method would struggle to solve the planning problem within a reasonable time. The SDDP algorithm scales better by the number of scenarios and stages, but not as much by the subproblem size. Thus, this is a significant issue to tackle to make this framework universally scalable, reducing the time required for the subproblem solution and increasing the speed of iterations by using better parallel schemes.

The optimality conditions of the algorithm are another vital point of discussion. Currently, the model has complete continuous recourse due to load shedding, which ensures a feasible continuous local variable value in every sub-problem. While convergence is guaranteed in the current SDDP model with finite support, as in this framework, optimality is not \citep{ papavasiliou2018application}. However, this framework remains an appropriate option for solving the large test case and might be the only framework that could provide a valuable solution at such a scale. It is also important to note that the performance of the SDDP algorithm varied with different initializations, and initializing the transition variables with zero resulted in a significantly longer solution time. Thus, an initialization scheme is required for performance consistency upon solving other test cases.

\subsection{Future work}

A fundamental improvement for this work is optimizing the computational performance of the proposed solution algorithm. The parallel scheme is not optimized and requires plenty of memory and communication time. Besides, the granularity potential within the framework has not been fully explored. A few schemes are proposed to enhance the current framework further. 

In the forward pass, instead of sharing a whole copy of the entire problem to different workers, the master process can hold all the information and then send forward pass samples to other workers to solve. In the backward pass, each subproblem has to be solved again with the generated cut for every scenario. In the current framework, this process is not parallelized. Parallelizing the forward and backward passes would require intricate programming and communication protocols to mitigate numerical errors and communication bottlenecks. Moreover, the existence of distributed gurobi \citep{gurobi} invokes the addition of a solution-level granularity where multiple cores can be assigned to each subproblem. This level of granularity is not witnessed in the literature using SDDP.

The efficacy of the SDDP algorithm for multistage stochastic problems is contingent on the number of scenarios and stages, which are the primary drivers behind its complexity. This issue is apparent in the AESO-144 test case, which maintained rather large subproblems. Implementing subproblem-level network and temporal decomposition, complemented by efficient parallelism, would enhance iterations’ computational performance and allow for a higher temporal resolution operationally. At every stage and scenario, the deterministic subproblem can be decomposed into smaller subproblems through temporal and network decomposition as done previously on static deterministic problems \citep{Fragkos_2021}. Finally, the number of variables for candidate generation and other technologies in the AESO-144 test case might be larger than necessary. In real-world planning cases, candidate technologies or new capacities may only apply to some buses and RoWs as additional constraints (such as land restrictions and regulatory, legal, and social concerns) may need to be considered.

With this diversity of solutions to the computation burden problem, the proposed model would exhibit enough efficiency and performance enhancement to include more factors aligned with the mission of achieving net-zero emissions. The envisioned additions to the proposed model include considering other greenhouse gases (such as $\text{SO}_6$ and $\text{N}_2\text{O}$), modeling more equipment degradation, modeling generator retirement, and incorporating resilient expansion against natural disasters (such as wildfires) which typically have an increasing rate as a significant consequence of the climate change. All can take precedence in future works and are a part of the ongoing investigation of the transition planning problem.

\section{Conclusion}
\label{section8}

This study inaugurates the Power System Transition Planning (PSTP) problem category, establishing a structured approach for guiding any jurisdictional power system toward a zero-emission future. The proposed model incorporates various modular planning factors, physical resources, and geospatial considerations, significantly enhancing its applicability across diverse system contexts. The formulation is based on a Multistage Dynamic Stochastic Programming structure, which introduces substantial computational complexity—addressed here through adopting the Stochastic Dual Dynamic Programming (SDDP) algorithm.

The paper presents a blueprint PSTP formulation, its linearization, and the implementation of the SDDP algorithm, including a breakdown of the post-decomposition subproblem structure. Extensive data sourcing and input processing procedures are detailed, followed by a series of numerical case studies. These evaluate model performance under various scenarios and validate that the SDDP algorithm can produce results comparable in accuracy to a monolithic MILP formulation on smaller systems. When deployed with parallel High-Performance Computing (HPC), the framework successfully scales to large, realistic systems within acceptable computational time frames.

Beyond its theoretical utility, the PSTP framework demonstrates substantial practical value. It offers a tractable and transparent tool to assist policymakers, system operators, and planning authorities in designing effective market structures, aligning stakeholder incentives, and crafting long-term strategies that support jurisdiction-specific efficiency and decarbonization targets.

\bibliographystyle{elsarticle-num} 
\bibliography{References}

\appendix

\section{Additional Tables}
\label{app:tables}

\begin{table}[!htbp]
\centering
\caption{Pan-regional Software Tools for Power System Planning}
\small
\label{tab:software-tools}
\resizebox{\textwidth}{!}{%
\begin{tabular}{l|lll}
\hline
\textbf{Software \textsuperscript{*}} & \textbf{Developer / Organization} & \textbf{Typical Users} & \textbf{Jurisdiction}   \\ \hline
PLEXOS \citep{PLEXOS2024} & Energy Exemplar & ERCOT, NGESO, CAISO, AEMO & Global  \\ 
PyPSA \citep{PyPSA2024} & Open-source (KIT, PyPSA-Eur) & Academia, research institutions & Global  \\ 
PowerFactory \citep{PowerFactory2024} & DIgSILENT GmbH & Utilities, TSOs, planners & Global  \\ 
PowerWorld \citep{PowerWorld2024} & PowerWorld Corporation & Utilities, ISOs/RTOs, academia & Global  \\ 
ReEDS \citep{ReEDS2023} & NREL & DOE, federal agencies, researchers & U.S. national  \\ 
SWITCH \citep{SWITCH2024} & UC Berkeley & Academia, governments, NGOs & U.S., Chile  \\ 
TYNDP \citep{TYNDP2022} & ENTSO-E & European TSOs, EU institutions & Europe    \\ 
PlanOS \citep{PlanOS2023}& GE Vernova & Utilities, industry planners & Global  \\ 
QuESt Planning \citep{QuESt2022} & Sandia National Laboratories & U.S. federal planners, researchers & U.S. \\ 
MATPOWER \citep{MATPOWER2024}& Cornell University & Academia, research groups & Global   \\ \hline
\end{tabular}%
}
{\scriptsize \textsuperscript{*} Includes only software with explicit network modeling and investment decisions}
\end{table}

\begin{table}[h!]
\centering
\caption{Comparison of long-term energy planning frameworks}
\label{tab:PlanPeriod}
\scriptsize
\resizebox{\textwidth}{!}{%
\begin{tabular}{@{}p{5cm}p{4cm}p{4cm}@{}}
\toprule
\textbf{Department} & \textbf{Planning Duration} & \textbf{Update Frequency} \\ \midrule
US Department of Energy \citep{USDOE_QTR} & 5–20 years & Every 4 years \\
TYNDP (ENTSO-E, Europe) \citep{ENTSOE_TYNDP} & 10–20 years & Every two years \\
AESO LTTP (Alberta, Canada) \citep{AESO_LTP} & 20 years & Every two years \\
ERCOT – LTSA (Texas, USA) \citep{ERCOT_LTSA} & Up to 15 years & Every two years \\
National Grid ESO Models (UK) \citep{NationalGrid_ESO} & 15–30 years & Annually \\
International Energy Agency \citep{IEA_LTEP} & 20–30 years & Annually \\
\bottomrule
\end{tabular}
}
\end{table}

\begin{table}[ht]  \caption{Existing Capacity and Load}
  \label{tab:ExCap}
  \centering
  \small
  \begin{tabular}{p{10pt}p{40pt}p{19pt}p{15pt}p{15pt}p{20pt}p{20pt}p{20pt}c}
    \toprule
     Bus & Name &  Coal & Gas & Bio & Hydro & Wind & Solar & Avg. Load \\
    \midrule
   1 & South     & -    & 14.0  & 4.00 & 9.31 & 118  &  55.5 & 85.2 \\ 
   2 & Calgary   & 58.0 & 123   & 1.00 & 17.0 & 7.00 &  3.15&116.1 \\ 
   3 & Central   & 51.0 & 42.0  & 3.00 & 41.0 & 17.0 &  1.69 & 123   \\
   4 & Edmonton  & 329  & 68.0  & 0.37 & -    & -    &-     & 151 \\ 
   5 & Northwest & 12.0 & 36.0  & 13.7 & -    & -    & -     & 80.6  \\ 
   6 & Northeast & -    & 142   & 8.90 & -    & -    & -     & 113 \\ 
    \bottomrule
  \end{tabular}
\end{table}

\begin{table}[ht]  \caption{Transmission RoW}
  \label{tab:TransRow}
  \centering
  \small
  \begin{tabular}{cccccc}
\toprule
Branch & From 	  & To 	      & Dist. (km) & Rating (p.u) & Existing  \\
\midrule
1 & Northwest & Northeast & 365.48	 & 0.2517	     & \checkmark  \\
2 & Northwest & Edmonton  & 389.9	 & 0.2517	     & \checkmark  \\
3 & Northeast & Edmonton  & 429.05	 & 0.5036	     & \checkmark  \\
4 & Edmonton  & Central	  & 139.38	 & 3.441	     & \checkmark  \\
5 & Central	  & Calgary	  & 139.08	 & 2.174	     & \checkmark  \\
6 & Calgary	  & South	  & 173.95	 & 0.8577	     & \checkmark  \\
7 & Northwest & Calgary	  & 621.42	 & 3.441		     & - \\
8 & Northwest & Central   & 497.54	 & 3.441		     & - \\
9 & Central	  & South	  & 296.01	 & 3.441		     & - \\
10 & Edmonton  & South	  & 450	     & 3.441	     & - \\
\bottomrule
  \end{tabular}
\end{table}

\end{document}